\documentclass[11pt]{article}

\let\fn\footnote
\renewcommand{\footnote}[1]{\linespread{1.1}\fn{#1}\linespread{1.29}}

\makeatletter\renewcommand{\section}{\@startsection
{section}{1}{\z@}{-3.5ex plus -1ex minus
    -.2ex}{2.3ex plus .2ex}{\bf }}
\makeatletter\renewcommand{\subsection}{\@startsection{subsection}{2}{\z@}{-3.25ex
plus -1ex minus
   -.2ex}{1.5ex plus .2ex}{\it }}
\makeatletter\renewcommand{\subsubsection}{\@startsection{subsubsection}{3}{-2.45ex}{-3.25ex
plus -1ex minus -.2ex}{1.5ex plus .2ex}{\it }}

\makeatletter \@addtoreset{equation}{section}

\renewenvironment{thebibliography}[1]
     {\baselineskip=16pt plus 2pt minus 1pt
      \section*{\large\refname
        \@mkboth{\MakeUppercase\refname}{\MakeUppercase\refname}}%
     \list{\@biblabel{\@arabic\c@enumiv}}%
           {\settowidth\labelwidth{\@biblabel{#1}}%
            \leftmargin\labelwidth
            \advance\leftmargin\labelsep
            \@openbib@code
            \usecounter{enumiv}%
            \let\p@enumiv\@empty
            \renewcommand\theenumiv{\@arabic\c@enumiv}}%
      \sloppy
      \clubpenalty4000
      \@clubpenalty \clubpenalty
      \widowpenalty4000%
      \sfcode`\.\@m}

\newcommand{\acknowledgements}{\section*{Acknowledgements}
\addcontentsline{toc}{section}{\hspace{0.6cm}{\bf Acknowledgements}}}

\usepackage{epsfig}
\usepackage{amsmath,amssymb}
\usepackage{amsfonts}
\usepackage{mathrsfs}
\usepackage{bbm}
\usepackage{bm}

\def\slasha#1{\setbox0=\hbox{$#1$}#1\hskip-\wd0\hbox to\wd0{\hss\sl/\/\hss}}

\def\periodb#1{\setbox0=\hbox{$#1$}#1\hskip-\wd0\hbox to\wd0{-}}

\newcommand{\binomr}[2]{\binom{\,#1\,}{\,#2\,}}

\newcommand{\unit}{\mathbbm{1}}   			

\newcommand{\CA}{\mathcal{A}}    			
\newcommand{\CAt}{\tilde{\mathcal{A}}}

\newcommand{\CF}{\mathcal{F}}

\newcommand{\CL}{\mathcal{L}}
\newcommand{\CM}{\mathcal{M}}
\newcommand{\CN}{\mathcal{N}}
\newcommand{\CO}{\mathcal{O}}

\newcommand{\CP}{\mathcal{P}}
\newcommand{\CPh}{\hat{\mathcal{P}}}

\newcommand{\CQh}{\hat{\mathcal{Q}}}

\newcommand{\CR}{\mathcal{R}}

\newcommand{\frm}{\mathfrak{m}}

\newcommand{\FR}{\mathbbm{R}}     			
\newcommand{\FC}{\mathbbm{C}}     			
\newcommand{\NN}{\mathbbm{N}}     			
\newcommand{\DD}{\mathbbm{D}}     			
\newcommand{\FF}{\mathbbm{F}}     			
\newcommand{\VV}{\mathbbm{V}}     			
\newcommand{\RZ}{\mathbbm{Z}}     			
\newcommand{\CPP}{{\mathbbm{C}P}}    			
\newcommand{\PP}{{\mathbbm{P}}}    			
\newcommand{\ah}{\hat{a}}
\newcommand{\Ah}{\hat{A}}
\newcommand{\fh}{\hat{f}}
\newcommand{\Ih}{\hat{I}}
\newcommand{\dd}{\mathrm{d}}     			
\newcommand{\dpar}{\partial}     			
\newcommand{\embd}{{\hookrightarrow}}     		
\newcommand{\di}{\mathrm{i}}     			

\newcommand{\bw}{{\bar{w}}}
\newcommand{\bz}{{\bar{z}}}

\newcommand{\eand}{{~~~\mbox{and}~~~}}     		
\newcommand{\ewith}{{~~~\mbox{with}~~~}}

\newcommand{\der}[1]{\frac{\dpar}{\dpar #1}}   		
\newcommand{\derr}[2]{\frac{\dpar #1}{\dpar #2}}   	
\newcommand{\tr}{\,\mathrm{tr}\,}     			
\newcommand{\ad}{\mathrm{ad}}     			

\newcommand{\au}{\mathfrak{u}}
\newcommand{\asu}{\mathfrak{su}}

\newcommand{\sU}{\mathsf{U}}     			

\newcommand{\sSU}{\mathsf{SU}}

\newcommand{\sS}{\mathsf{S}}

\newcommand{\sEnd}{\mathsf{End}\,}
\newcommand{\sSpecM}{\mathsf{SpecM}\,}
\newcommand{\vac}{|0\rangle}
\newcommand{\cav}{\langle 0|}
\newcommand{\remark}[1]{}     				
\def\tyng(#1){\hbox{\tiny$\yng(#1)$}}			

\begin{document}
\begin{titlepage}
\begin{flushright}
  hep-th/0612173\\
  DIAS-STP-06-24
\end{flushright}
\vskip 2.0cm
\begin{center}
{\LARGE \bf Fuzzy Toric Geometries}
\vskip 1.5cm
{\Large Christian S{\"a}mann}
\setcounter{footnote}{0}
\renewcommand{\thefootnote}{\arabic{thefootnote}}
\vskip 1cm
{\em School of Theoretical Physics\\
Dublin Institute for Advanced Studies\\
10 Burlington Road, Dublin 4, Ireland}\\[5mm]
{Email: {\ttfamily csamann@stp.dias.ie}} \vskip
1.1cm
\end{center}
\vskip 1.0cm
\begin{center}
{\bf Abstract}
\end{center}
\begin{quote}
We describe a construction of fuzzy spaces which approximate projective toric varieties. The construction uses the canonical embedding of such varieties into a complex projective space: The algebra of fuzzy functions on a toric variety is obtained by a restriction of the fuzzy algebra of functions on the complex projective space appearing in the embedding. We give several explicit examples for this construction; in particular, we present fuzzy weighted projective spaces as well as fuzzy Hirzebruch and del Pezzo surfaces. As our construction is actually suited for arbitrary subvarieties of complex projective spaces, one can easily obtain large classes of fuzzy Calabi-Yau manifolds and we comment on fuzzy K3 surfaces and fuzzy quintic three-folds. Besides enlarging the number of available fuzzy spaces significantly, we show that the fuzzification of a projective toric variety amounts to a quantization of its toric base.
\end{quote}
\end{titlepage}

\section{Introduction}

Consider two complex affine varieties $X\subset \FC^m$ and $Y\subset \FC^n$ together with their coordinate rings $R(X)\subset\FC[x_1,\ldots ,x_m]$ and $R(Y)\subset\FC[y_1,\ldots ,y_n]$. A well-known theorem in algebraic geometry states that there is a one-to-one correspondence between morphisms $f: X\rightarrow Y$ and $\FC$-algebra homomorphisms $f^*:R(Y)\rightarrow R(X)$. This is easy to see: First of all, $f^*$ uniquely pulls back polynomials on $X$ to $Y$. Inversely, by $f=(f_1,\ldots ,f_n)$ with $f_i=f^*(y_i)$, we have a canonical morphism for every $\FC$-algebra homomorphism $f^*$. Due to this theorem, points in $X$ correspond to maximal ideals in $R(X)$ and we can identify $X$ with the set of maximal ideals, the {\em maximal spectrum}, in $R(X)$: $X=\sSpecM R(X)$.

In noncommutative geometry, one makes use of this relation between algebraic varieties and their coordinate ring: Instead of quantizing the space itself, one quantizes the algebra of functions living on the space by truncating the algebra and/or deforming the product structure. If the thus obtained algebras are isomorphic to finite dimensional matrix algebras, which is the case e.g.\ for symplectic coset spaces, the resulting noncommutative geometries are called {\em fuzzy}.

These fuzzy spaces are interesting essentially for two reasons: The first one is that the fuzzy framework provides a nice way of regularizing quantum field theories on compact Riemannian spaces without breaking spacetime symmetries. It is therefore considered a useful alternative to the lattice approach. Second, fuzzy spaces arise naturally in string theory when one considers D-brane configurations in certain nontrivial background fields, see e.g.\ \cite{Myers:1999ps}. 

The fuzzy spaces studied in the literature so far are the fuzzy sphere \cite{Hoppe:Diss} and orbifolds thereof \cite{Martin:2004dm}, the fuzzy disc \cite{Lizzi:2003ru}, the fuzzy complex projective spaces \cite{Balachandran:2001dd} and deformations thereof \cite{Ramgoolam:2001zx}, fuzzy tori \cite{Hoppe:1988qt}, the fuzzy supersphere \cite{Grosse:9804013}, and fuzzy Gra{\ss}mannians as well as fuzzy flag manifolds together with their superextensions \cite{Dolan:2001mi,Ivanov:2003qq,Murray:2006pi}, see also \cite{Trivedi:2000mq}. This set of spaces is still very limited, and hence it is desirable to find further examples of fuzzy spaces.

In particular, compact spaces appearing in string theory's compactification scenarios are of major interest and one is therefore naturally led to examine projective toric varieties. From an algebraic geometric point of view, these spaces are also the obvious next step after the fuzzification of the classical projective spaces.

The key property used in our approach to rendering projective toric varieties fuzzy is that these spaces have (by definition) a natural interpretation as subvarieties of complex projective spaces. Recall that given a variety $X$, one can exclude a subset $S$ of $X$ by restricting to those polynomials in the coordinate ring $R(X)$ which vanish on $S$:
\begin{equation}
R(X\backslash S)\ =\ \{ f\in R(X)\,|\,f(S)=0 \}~.
\end{equation}
To obtain the subset $S$ itself, one factors out the ideal $I$ generated by the elements in $R(X\backslash S)$ from the full coordinate ring $R(X)$: 
\begin{equation}
R(S)\ =\ R(X)/I~. 
\end{equation}
The same holds for projective varieties and their projective subsets after restricting to homogeneous ideals, and this will yield a natural quantization procedure for such spaces. 

In particular, we will show that for a toric variety, this quantization procedure corresponds to quantizing the underlying toric base. Besides establishing this result, the purpose of this paper is to provide a starting point for further studies of fuzzy toric geometries. Some interesting future directions are mentioned in the conclusions. 

The outline of our presentation is as follows: In section 2, we begin with a concise review of the description of fuzzy $\CPP^n$ to establish our notation. Our construction is presented in detail using the simple example of the Veronese surface in section 3. Section 4 deals with some new features arising in the case of weighted projective spaces. After briefly reviewing the basics of toric geometry in section 5, we present the general algorithm for constructing fuzzy projective toric varieties in section 6. Section 7 is devoted to a number of interesting examples of our construction, and we summarize our results in section 8.

\section{Fuzzy complex projective spaces}

\subsection{Matrix algebra on $\CPP^n_F$}

The prototype of all fuzzy spaces is certainly the fuzzy complex projective space $\CPP^n_F$, as the constructions of almost all\footnote{i.e.\ except for the fuzzy tori} other fuzzy geometries are derived from it; fuzzy toric varieties will be no exception. The reason for this prominent r{\^o}le is the fact that $\CPP^n$ is the space $\sU(n)/\big(\sU(1)\times \sU(n-1)\big)\cong \sSU(n)/\sS\big(\sU(1)\times \sU(n-1)\big)$ and it is this coset description, which allows for a particularly nice quantization prescription; see \cite{Murray:2006pi} for a detailed discussion of the quantization of such coset spaces.

The first aspect of describing a fuzzy geometry is to give a sequence of matrix algebras approximating the algebra of functions on this space. Consider the space $\FC^{n+1}$ with its polynomial ring $R[\FC^{n+1}]=\FC[w_0,\ldots ,w_n]$ and its restrictions to homogeneous polynomials of degree $L$, $R_L$. Real analytic functions on $\FC^{n+1}$ can be Taylor expanded in terms of elements of $R_L\otimes R^*_K$ plus their complex conjugate, where $L$ and $K$ run over the natural numbers. If we factor out the ideal $I$ generated by $w_i\bw_i-1$ from $R\otimes R^*$, we descend to functions on $S^{2n+1}$. These are therefore expanded in terms of (real combinations of) the normalized monomials
\begin{equation}\label{continuumbasis}
\frac{w_{i_1}\ldots w_{i_L}\bw_{j_1}\ldots \bw_{j_K}}{r^{L+K}}\ewith r\ =\ \sqrt{w_i\bw_i}~,
\end{equation}
which are (normalized) elements of $R_L\otimes R^*_K$. The short exact sequence
\begin{equation}
1\ \longrightarrow\  \sU(1) \ \longrightarrow\  S^{2n+1} \ \longrightarrow\  \CPP^n \ \longrightarrow\  1
\end{equation}
furthermore tells us that the real analytic functions on $\CPP^n$ are the functions on $S^{2n+1}$, which are invariant under a $\sU(1)$ action. This action can be taken to be the multiplication of the vector $w$ by a phase, and real analytic functions on $\CPP^n$ are thus expanded in terms of elements of $\CM_L:=R_L\otimes R_L^*$, where $L$ runs over the natural numbers. The $w_i$ then find their usual interpretation as homogeneous coordinates on $\CPP^n$. 

Note that since we factored out the ideal $I=\{w_i\bw_i-1\}$, any element of $\CM_{L-1}$ can be written as a contraction of an element of $\CM_{L}$ :
\begin{equation}\label{identcontinuum}
\frac{w_{i_1}\ldots w_{i_{L-1}}\bw_{j_1}\ldots \bw_{j_{L-1}}w_k\bw_k}{r^{2L}}\ =\ \frac{w_{i_1}\ldots w_{i_{L-1}}\bw_{j_1}\ldots \bw_{j_{L-1}}}{r^{2(L-1)}}~.
\end{equation}

The noncommutative picture arises by replacing complex coordinates with the creation and annihilation operators of $n+1$ harmonic oscillators. On noncommutative $\FC^{n+1}$, we have
\begin{equation}
 w^i \ \rightarrow\  \ah_i\eand \bw^i\ \rightarrow\  \ah_i^\dagger \ewith [\ah_i,\ah^\dagger_j]\ =\ \delta_{ij}~.
\end{equation}
The elements of $R_L\otimes R_K^*$ naturally become operators, and the ordinary operator product yields an infinite dimensional algebra. Note that by rescaling the creation and annihilation operators of the individual oscillators and subsequent rotations, one can arrive at arbitrary deformation tensors $\Theta^{\mu\nu}$ on noncommutative $\FR^{2n+2}\cong \FC^{n+1}$.

Noncommutative $S^{2n+1}$ in operator language is obtained by factoring out the bi-ideal corresponding to $I$. For this, we define the operators
\begin{equation}
 \hat{b}_i\ :=\ \ah_i\frac{1}{\sqrt{\hat{N}}}\eand
\hat{b}^\dagger_i\ :=\ \frac{1}{\sqrt{\hat{N}}} \ah^\dagger_i~,
\end{equation}
where $\hat{N}=\ah^\dagger_i\ah_i$ is the usual number operator. Note that $\hat{b}_i^\dagger\hat{b}_i-\unit=0$, and therefore switching to the operators $\hat{b},\hat{b}^\dagger$ corresponds to factoring out the appropriate ideal.

Observe that we can conveniently rewrite the basis elements of the operator algebra on noncommutative $\FC^{n+1}$ in the following way: we introduce normal ordering of all creation and annihilation operators, and insert a ``double vacuum'' between the two species obtaining operators of the form\footnote{See \cite{Dolan:2006tx,Murray:2006pi} for a more detailed discussion of this point.}
\begin{equation}
\ah_{i_1}^\dagger\ldots \ah^\dagger_{i_L}\vac\cav \ah_{j_1}\ldots \ah_{j_K}\in \CA_{L,K}~.
\end{equation}

As noted above, we descend to functions on $\CPP^n$ by considering polynomials $\CM_L$, i.e.\ by fixing $L=K$. In this case, the number operator is just a constant, and the detour via the operators $\hat{b},\hat{b}^\dagger$ on $S^{2n+1}$ is not necessary. We can implement this restriction by projecting out all monomials of degree $d\neq L$,
\begin{equation}\label{proj1}
\CA_L\ :=\ \CPh_L\left(\bigoplus_{M,N}\CA_{M,N}\right)\CPh_L \ewith
\CPh_L\ :=\ \frac{1}{\CN_{i_1\ldots i_L}}\ah^\dagger_{i_1}\ldots \ah^\dagger_{i_L}\vac\cav\ah_{i_1}\ldots \ah_{i_L}~,
\end{equation}
where $\CN_{...}$ are normalization constants ensuring that $\CPh_L$ acts like the identity operator on $\CA_{L,L}$. The resulting monomials
\begin{equation}\label{operatorbasis}
\ah^\dagger_{i_1}\ldots \ah^\dagger_{i_L}\vac\cav \ah_{j_1}\ldots \ah_{j_L}
\end{equation}
span the fuzzy algebra of functions truncated at level $L$, $\CA_L=\CA_{L,L}=:\CR_L\otimes\CR_L^*$, which defines the fuzzy complex projective space $\CPP^n_F$. Here, $\CR_L$ denotes the restriction of the Fock space of $n+1$ harmonic oscillators to its $L$-particle Hilbert subspace. The operators \eqref{operatorbasis} evidently form a finite dimensional algebra in which multiplication is defined as the ordinary operator product. Note that this is in fact the case for arbitrary projectors $\CPh_L$ of finite rank, and we will make use of this observation in the quantization of subvarieties of $\CPP^n$. For our choice of $\CPh_L$, the algebra $\CA_L$ is isomorphic to the algebra of square matrices $\sEnd(\FC^{d_L})$, where $d_L=\frac{(N+L)!}{N!L!}$ \cite{Balachandran:2001dd}. We interpret elements of $\CA_L$ as approximations to functions since after taking the limit $L\rightarrow \infty$ in an appropriate way, $\CA_L$ tends to the ordinary algebra of real functions on $\CPP^n$ \cite{Balachandran:2001dd}.

\subsection{Star product on $\CPP^n_F$}

Besides the operator approach, we can describe the algebra of fuzzy functions on $\CPP^n_F$ by restricting to the set of monomials of order $L$ in both the homogeneous coordinates and their complex conjugates, which we denoted by $\CM_L$. Consider the coherent states truncated at level $L$,
\begin{equation}\label{coherentstates}
|w,L\rangle\ :=\ \frac{1}{\sqrt{L!}}\left(w_i\ah^\dagger_i\right)^L \vac\ =\ \left(w_i\hat{b}^\dagger_i\right)^L \vac~.
\end{equation}
Given a homogeneous polynomial $p\in \FC[t_0,\ldots ,t_n]$ of degree $L$, we have the formula
\begin{equation}\label{Fformula}
 p(\ah_0,\ldots ,\ah_{n})|w,L\rangle\ =\ p(w_0,\ldots ,w_{n})|0\rangle~.
\end{equation}
The coherent states \eqref{coherentstates} allow us to introduce a natural map from operators in $\CA_L$ to functions on $\CPP^n$:
\begin{equation}\label{coherentstatemap}
f(w)\ =\ \CF_L(\hat{f})\ :=\ \tr\left(\hat{\rho}(w,L)\hat{f}\right)\ewith \hat{\rho}(w,L)\ =\ |w,L\rangle{}\langle w,L|~.
\end{equation}
Due to \eqref{Fformula}, $\CF_L$ maps the basis elements \eqref{operatorbasis} to the monomials \eqref{continuumbasis}, up to an interchange of the indices $i_k$ and $j_k$. Therefore, this map is bijective and together with \eqref{identcontinuum}, this motivates the inclusion of $\CA_{L-1}$ into $\CA_L$ via the identification\footnote{This identification is very natural from a group theoretic point of view, see e.g.\ \cite{Murray:2006pi}. Note that $\CR_L$ forms a representation space of $\asu(n+1)$ for the representation corresponding to the Dynkin labels $(L,0,\ldots ,0)$, while $\CR_L^*$ is related to a representation $(0,\ldots ,0,L)$. The tensor product of these representations contains the tensor products of such representations for any lower value of $L$.} 
\begin{equation}\label{identfuzzy}
\ah^\dagger_{i_1}\ldots \ah^\dagger_{i_{L-1}}\vac\cav \ah_{j_1}\ldots \ah_{j_{L-1}}~~\sim~~
\ah^\dagger_{i_1}\ldots \ah^\dagger_{i_{L-1}}\ah^\dagger_k\vac\cav \ah_k \ah_{j_1}\ldots \ah_{j_{L-1}}~.
\end{equation}
This inclusion is important, as it shows that each matrix algebra at level $L$ approximates the algebra of functions on $\CPP^n$ at least as well as a matrix algebra at lower levels.

Moreover, the map $\CF_L$ induces a deformed or star product on $\CM_L$ \cite{Balachandran:2001dd} via
\begin{equation}
(f\star g)(w)\ :=\ \CF_L(\hat{f}\hat{g})\ =\ \tr\left(\hat{\rho}(w,L)\hat{f}\hat{g}\right)~,
\end{equation}
where $f=\CF_L(\hat{f})$ and $g=\CF_L(\hat{g})$ are the functions corresponding to the operators $\hat{f}$ and $\hat{g}$. Together with this deformed product, the set $\CM_L$ forms indeed an algebra, which we denote by $(\CM_L,\star)$. In the limit $L\rightarrow \infty$, it is possible to show that the star product goes over into the ordinary product between real analytic functions on $\CPP^n$ \cite{Balachandran:2001dd}. In its simplest form, the deformed product reads as \cite{Kurkcuoglu:2006iw}
\begin{equation}
f\star g\ =\ \mu\left[\left(\frac{1}{L!}\der{w_{i_1}}\ldots \der{w_{i_L}}\otimes \frac{1}{L!}\der{\bw_{i_1}}\ldots \der{\bw_{i_L}}\right)(f\otimes g)\right]\ewith \mu[a\otimes b]\ =\ a\cdot b~.
\end{equation}

\subsection{The Riemannian geometry of $\CPP^n_F$}

To define field theories on fuzzy $\CPP^n$, we require some additional structure corresponding to the metric in the continuum. Even for a scalar field theory, it is necessary to define a Laplace operator; further differential operators are needed in more general theories. On ordinary manifolds embedded into flat space, we can simply pull back the flat metric to the embedded space. Furthermore, if the ambient space is a K{\"a}hler manifold and the embedding is holomorphic (and thus the embedded manifold is a complex submanifold), the submanifold is also K{\"a}hler. 

In the case of $\CPP^n$, we can use the obvious embedding $\CPP^n\embd\FC^{n+1}$ and the pull back of the canonical K{\"a}hler metric on $\FC^{n+1}$ will give rise to the K{\"a}hler metric known as the Fubini-Study metric. Equivalently, we could use the natural embedding of $\CPP^n$ into $\FR^{(n+1)^2}$. This embedding is a result of the correspondence between a point on $\CPP^n$ and a rank 1 projector on $\FC^{n+1}$: Such a projector can be expanded in terms of the $(n+1)^2-1$ Gell-Mann matrices of $\asu(n+1)$, $\lambda^a_{ij}$, $a=1,\ldots ,(n+1)^2-1$, together with the unit matrix $\lambda^0_{ij}=\delta_{ij}$. The embedding $\CPP^n\embd \FR^{(n+1)^2}$ is given explicitly by $x^{\ah}=\bw^i\lambda^{\ah}_{ij}w^j$, $\ah=0,\ldots ,(n+1)^2-1$, and the Euclidean metric on $\FR^{(n+1)^2}$ together with this space's canonical complex structure induces again the usual Fubini-Study metric. The derivatives spanning the tangent space to $\CPP^n$ in terms of the coordinates on $\FR^{(n+1)^2}$ read as
\begin{equation}
\CL_a\ =\ -\di f_{ab}{}^c x^b\der{x^c}~,
\end{equation}
where $f_{ab}{}^c$ are the structure constants of $\asu(n+1)$. 

By demanding that the derivatives act on functions in $(\CM_L,\star)$ as they would in the continuum, we obtain induced derivations in $\CA_L$ from compatibility with \eqref{coherentstatemap} \cite{Balachandran:2001dd}. Explicitly, the action of the derivatives $\CL_a$ is mapped to the adjoint action of the generators of $\asu(n+1)$ in the Schwinger construction:
\begin{equation}
\CL_a \ \rightarrow\  [\hat{L}^a,\,\cdot\,]\ :=\  [\ah^\dagger_i\lambda_{ij}^a \ah_j,\,\cdot\,]~.
\end{equation}
The Laplacian $\delta^{ab}\CL_a\CL_b$ is then naturally mapped to the second order Casimir operator acting in the adjoint,
\begin{equation}
\Delta \ \rightarrow\ \hat{\Delta}\ :=\ \ad(C_2)\,\cdot\,=\delta^{ab} [\hat{L}^a,[\hat{L}^b,\,\cdot\,]]~.
\end{equation}
This fixes the actual geometry of $\CPP^n_F$ sufficiently for our purposes.

\section{The fuzzy Veronese surface}

Our approach to fuzzy toric geometries is based on the possibility of considering the projective toric varieties as subvarieties of complex projective spaces. Let us start in this section with a detailed discussion of a particularly simple example of quantizing a subvariety of $\CPP^n$: the fuzzy Veronese surface $\VV_{2,2}$.

\subsection{Embedding of the Veronese surface in $\CPP^5$}

The {\em Veronese surface} $\VV_{2,2}$ is defined as an embedding $\nu_{2,2}$ of $\CPP^2$ in $\CPP^5$. In homogeneous coordinates, this map reads explicitly as
\begin{equation}\label{Veronese2}
\nu_{2,2}: (z_0,z_1,z_2)\ \mapsto\  (z_0^2,z_1^2,z_2^2,z_0z_1,z_0z_2,z_1z_2)\ =\ (w_i)~.
\end{equation}
It is straightforward to generalize $\nu_{2,2}$ by starting from an arbitrary complex projective space $\CPP^m$ and using homogeneous polynomials of arbitrary degree $d$. One thus arrives at the Veronese variety of degree $d$, which yields an embedding $\nu_{m,d}:\CPP^m\embd \CPP^n$ with $n=\binomr{m+d\,}{d}-1$.

Closely related to this picture is the so-called {\em Segre embedding}, which defines a map $\mu_{mn}:\CPP^m\times \CPP^n\embd \CPP^{(m+1)(n+1)-1}$ and thus proves that the product of two complex projective spaces is a projective variety. In terms of homogeneous coordinates on the involved spaces, the Segre embedding reads as
\begin{equation}
\mu_{mn}:(x_0,\ldots ,x_n,y_0,\ldots ,y_m)\ \mapsto\  (x_0y_0,x_0y_1,\ldots ,x_my_n)~.
\end{equation}

Another such embedding is the {\em Pl{\"u}cker embedding} giving rise to Gra{\ss}mannian manifolds, whose fuzzification is discussed in detail in \cite{Murray:2006pi}.

For our further discussion, we need explicitly the homogeneous polynomials generating the ideal $I$, which we factor out from the homogeneous coordinate ring on $\CPP^5$ to obtain the corresponding coordinate ring on $\VV_{2,2}$. From \eqref{Veronese2}, we read off the six independent hyperquadric conditions
\begin{equation}\label{embedVeronese}
\begin{aligned}
&I_1\ :=\ w_0w_1-w_3^2\ =\ 0~,~~~&&I_2\ :=\ w_0w_2-w_4^2\ =\ 0~,\\
&I_3\ :=\ w_1w_2-w_5^2\ =\ 0~,~~~&&I_4\ :=\ w_3w_4-w_0w_5\ =\ 0~,\\
&I_5\ :=\ w_3w_5-w_1w_4\ =\ 0~,&~~~&I_6\ :=\ w_4w_5-w_2w_3\ =\ 0~.
\end{aligned}
\end{equation}
Such relations are easily found for general embeddings, but we need to prove in each case that we have got indeed the full set of polynomials generating the appropriate ideal. This proof is rather straightforward for the Veronese surfaces $\VV_{2,2}$ using the following picture. Identify each coordinate $w_i$ with a vector in three-dimensional space $\RZ^3$, where the entries correspond to the powers of $z_\alpha$ in $w_i(z_\alpha)$: 
\begin{equation*}
 w_0: \left(\begin{array}{c}2\\0\\0\end{array}\right)~,~~w_1: \left(\begin{array}{c}0\\2\\0\end{array}\right)~,~~w_2: \left(\begin{array}{c}0\\0\\2\end{array}\right)~,~~w_3: \left(\begin{array}{c}1\\1\\0\end{array}\right)~,~~w_4: \left(\begin{array}{c}1\\0\\1\end{array}\right)~,~~w_5: \left(\begin{array}{c}0\\1\\1\end{array}\right)~.
\end{equation*}
An identity is thus given by two distinct paths starting from the origin and ending at the same point in\footnote{Obviously, only the completely positive octant of this space is of interest.} $\RZ^3$. To avoid trivial identities, we demand that our path is normal ordered and thus all powers of $w_i$ come before those of $w_j$ for $i<j$. Common parts of the paths can obviously be erased. The same holds for two parts of the paths, which are identical up to the nontrivial identities $I_i$. The set of $I_i$ is complete if for any pair of paths, these operations yield no remainder.

Consider now two arbitrary such paths. We start by using the identities $I_i$ to get as many components of $w_0$ and $w_5$ as possible in the paths. Then we erase common powers of $w_0$ and $w_5$, and besides further powers of these two coordinates, the remaining paths are of two types: they are either built of $w_1$s and at most one $w_3$ or they consist of $w_2$s and at most one $w_4$. From rather trivial considerations, we can make the following statements: Two paths of the same type have to be identical to have a common endpoint, even if powers of $w_0$ and $w_5$ are assigned to either of the paths. Two paths of different type can never arrive at the same endpoint. Altogether, we can conclude that any identity reduces completely using $I_1,...,I_6$ and we thus have the complete set of homogeneous polynomials generating the ideal $I$.

Although this approach seems a little complicated for dealing with the Veronese surface, it trivially generalizes to arbitrary embeddings of projective varieties into complex projective spaces.

\subsection{Towards the fuzzy matrix algebra}

We observe that all the functions on the Veronese surface embedded in $\CPP^5$ are obtained by restricting a function on the ambient space. We thus have $R(\VV_{2,2})=R(\CPP^5)/I$, where $I$ is the ideal generated by $I_1,...,I_6$ as defined in \eqref{embedVeronese}. 

For quantization, we should therefore start from the polynomials $\tilde{\CM}_L=(R(\CPP^5)/I)_L\otimes (R(\CPP^5)/I)^*_L$. We can easily factor out the ideal by replacing all equivalent elements of $R(\CPP^5)_L$ by their average. The reason for averaging rather than picking one representative will become clearer when we will define a Laplace operator in section 3.5. Explicitly, we substitute the $w_i$ in the monomials by their expressions $w_i(z_\alpha)$ in terms of coordinates $z_\alpha$ on $\CPP^2$ and then averaging over all those monomials in the $w_i$ which yield the same expressions in the $z_\alpha$. For example, $(R(\CPP^5)/I)_2$ is spanned by the monomials 
\begin{equation}\label{spancontinuous}
\begin{aligned}
&w_0w_0~,~~~w_1w_1~,~~~w_2w_2~,~~~w_0w_1+w_3^2~,~~~w_0w_2+w_4^2~,\\
&w_1w_2+w_5^2~,~~~
w_0w_3~,~~~w_0w_4~,~~~w_0w_5+w_3w_4~,~~~w_1w_3~,\\
&w_1w_4+w_3w_5~,~~~w_1w_5~,~~~w_4w_5+w_2w_3~,~~~w_2w_4~,~~~w_2w_5~.
\end{aligned}
\end{equation}

The above considerations translate straightforwardly to the operator picture. By replacing homogeneous coordinates with creation and annihilation operators, we arrive at the operators $\CAt_L=\tilde{\CR}_L\otimes \tilde{\CR}_L^*$ forming the fuzzy algebra of functions. We can equivalently start from the coherent state map $\tilde{\CF}_L$ obtained from the truncated coherent states
\begin{equation}
|w,L\rangle\ =\ \frac{1}{\sqrt{L!}}\left(w_i(z_\alpha) \ah^\dagger_i\right)^L \vac~,
\end{equation}
where the $w_i$ are again written in terms of the coordinates $z_\alpha$ on $\CPP^2$. We have to replace all operators mapping to the same function under $\tilde{\CF}_L$ by their average. Both prescriptions yield, e.g., that $\tilde{\CR}_2$ is spanned by the states\footnote{There is a choice of inserting factors of $\frac{1}{n!}$ in front of operators $(\hat{a}^\dagger_i)^n$ for normalization purposes. See the discussion in the next section for details.}
\begin{equation}\label{spanfuzzy}
\begin{aligned}
&\ah^\dagger_0\ah^\dagger_0\vac~,~~~\ah^\dagger_1\ah^\dagger_1\vac~,~~~\ah^\dagger_2\ah^\dagger_2\vac~,~~~(\ah^\dagger_0\ah^\dagger_1+(\ah^\dagger_3)^2)\vac~,~~~(\ah^\dagger_0\ah^\dagger_2+(\ah^\dagger_4)^2)\vac~,\\
&(\ah^\dagger_1\ah^\dagger_2+(\ah^\dagger_5)^2)\vac~,~~~
\ah^\dagger_0\ah^\dagger_3\vac~,~~~\ah^\dagger_0\ah^\dagger_4\vac~,~~~(\ah^\dagger_0\ah^\dagger_5+\ah^\dagger_3\ah^\dagger_4)\vac~,~~~\ah^\dagger_1\ah^\dagger_3\vac~,\\
&(\ah^\dagger_1\ah^\dagger_4+\ah^\dagger_3\ah^\dagger_5)\vac~,~~~\ah^\dagger_1\ah^\dagger_5\vac~,~~~(\ah^\dagger_4\ah^\dagger_5+\ah^\dagger_2\ah^\dagger_3)\vac~,~~~\ah^\dagger_2\ah^\dagger_4\vac~,~~~\ah^\dagger_2\ah^\dagger_5\vac~.
\end{aligned}
\end{equation}
It is rather obvious that the operator product closes and that $\CAt_L$ forms an algebra, since $(\tilde{\CR}_L\otimes \tilde{\CR}_L^*)\cdot(\tilde{\CR}_L\otimes \tilde{\CR}_L^*)= (\tilde{\CR}_L\otimes \tilde{\CR}_L^*)$. Together with the star product induced from the coherent state map $\CF_L$ on $\CPP^5$, the algebras $\CAt_L$ and $(\tilde{\CM}_L,\star)$ are again isomorphic.

In the case of the Segre embedding, a similar construction shows that we can write the algebra of functions on the product space $\CPP^1_F\times \CPP^1_F$ at levels $(L,L)$ as the algebra of functions on $\CPP^3_F$ at level $L$ after appropriately factoring out an ideal. 

\subsection{The projection $\CA_{L}\rightarrow \CAt_L$}

Although we gave reasonable motivation for the averaging procedure from the continuum description, it is desirable to have a more explicit construction, particularly for higher values of $L$. Analogously to the construction of the fuzzy algebra of functions on $\CPP^n$ from the noncommutative algebra on $\FC^{n+1}$, we are looking for an operator $\CPh_L$, which projects from $\CA_L$ down to $\CAt_L:=\CPh_L\CA_L\CPh_L$. As usual for projectors, we demand that $\CPh_L^2=\CPh_L$ and $\CPh_L^\dagger=\CPh_L$. 

Recall that under quantization, an equation $I_i(w,\bw)=0$ should turn into an operator equation $\hat{I}_i(\ah^\dagger,\ah)|\psi\rangle=0$, where $|\psi\rangle$ is an arbitrary state in the relevant Hilbert space. We are thus led to introduce the six operators
\begin{equation}\label{3.7}
\begin{aligned}
\Ih_1\ =\ &\ah^\dagger_0\ah^\dagger_1-(\ah^\dagger_3)^2~,~~~&\Ih_2&\ =\ \ah^\dagger_0\ah^\dagger_2+(\ah^\dagger_4)^2~,~~~&\Ih_3&\ =\ \ah^\dagger_1\ah^\dagger_2+(\ah^\dagger_5)^2~,\\
\Ih_4\ =\ &\ah^\dagger_0\ah^\dagger_5+\ah^\dagger_3\ah^\dagger_4~,~~~&\Ih_5&\ =\ \ah^\dagger_1\ah^\dagger_4+\ah^\dagger_3\ah^\dagger_5~,~~~&\Ih_6&\ =\ \ah^\dagger_4\ah^\dagger_5+\ah^\dagger_2\ah^\dagger_3~,
\end{aligned}
\end{equation}
and demand that\footnote{$\fh\Ih_i^\dagger=0$ follows by complex conjugation for real functions $f$, i.e.\ hermitian operators $\fh$.} $\Ih_i\fh=0$ for all $\fh\in\CAt_L$. A projector which guarantees e.g.\ $\Ih_1\fh=0$ is given by
\begin{equation}\label{defP}
\CPh_{1;L}\ =\ (\unit_L-\CQh_{I_1;L})~,
\end{equation}
where
\begin{equation}\label{defQ}
\begin{aligned}
\unit_L&\ :=\ \frac{1}{\CN_{i_1\ldots i_L}}\ah^\dagger_{i_1}\ldots \ah^\dagger_{i_{L}}\vac\cav \ah_{i_1}\ldots \ah_{i_{L}}~,\\
\CQh_{I_1;L}&\ :=\ \frac{1}{\CN_{i_1\ldots i_{L-2}}}\Ih_1^\dagger\ah^\dagger_{i_1}\ldots \ah^\dagger_{i_{L-2}}\vac\cav \ah_{i_1}\ldots \ah_{i_{L-2}}\Ih_1 ~. 
\end{aligned}
\end{equation}
Here, the $\CN_{\ldots }$ are the obvious normalization constants ensuring $\unit^2_L=\unit_L$ and $\CQh_{I_1;L}^2=\CQh_{I_1;L}$, cf.\ \eqref{proj1}. As one can easily verify, all elements $\fh$ of the algebra $\CPh_{1;L}\CA_L\CPh_{1;L}$ satisfy the operator equation $\Ih_1\fh=0$.

To guarantee $\Ih_i\fh=0$ for all $i$, we need to build a projector $\CPh$ from all $\CPh_{i;L}$. Note that at level $L>2$, the projectors $\CPh_i$ are no longer orthogonal. This implies that the na{\"i}ve product $\CPh_{i;L}\CPh_{j;L}\neq \CPh_{j;L}\CPh_{i;L}$ and $\CPh_{i;L}\CPh_{j;L}$ is not a projector. One therefore has to find the unique projector $\CPh_{ij;L}$ whose image is the intersection of the images of $\CPh_{i;L}$ and $\CPh_{j;L}$. Putting all together, we arrive at 
\begin{equation}
\CAt_L\ :=\ \CPh_{123456;L}\CA_L\CPh_{123456;L}~.
\end{equation}
The explicit form of $\CPh_{123456;L}$ can be easily calculated but since we are more interested in the principles of quantizing subvarieties, we refrain from going into further details. We will, however, give detailed expressions for the space $W\CPP^2(1,1,2)$ in section 4.2. In the following, we will shorten our notation and use $\CPh_L:=\CPh_{123456;L}$.

Note that the algebra $\CAt_L$ obtained via the projector method is obviously identical to the algebra we defined by the averaging procedure with one minor exception: as $\cav \hat{a}_0\hat{a}_1 \hat{a}^\dagger_0\hat{a}^\dagger_1\vac=\tfrac{1}{2}\cav \hat{a}_3 \hat{a}_3 \hat{a}^\dagger_3\hat{a}^\dagger_3\vac$, one has to insert factors of $\frac{1}{n!}$ in front of $(\hat{a}^\dagger)^n$ either into the projectors $\hat{\CP}_L$ or into the states from which we construct the algebra of functions. We choose to preserve the quantization prescriptions for constraints outlined above \eqref{3.7} and thus use the latter convention in the following. 

Later on, we will see that the quantization of toric varieties can be performed without explicitly using the above projectors, as the toric bases underlying the definition of these varieties provide a more direct prescription of how to factor out the ideal.

Due to the isomorphy between $(R(\CPP^5)/I)_L$ and $\tilde{\CR}_L$, we can calculate the dimensions of the matrix algebras $\tilde{\CA}_L$: Using a computer algebra program, we produce\footnote{This is done by taking the product of all generators $I_i$ with $L-2$ arbitrary coordinates and eliminating triangle identities of the type $0=i_1-i_2=i_1-i_3=i_2-i_3$.} a set of polynomials which form a basis for $I_L$, the set of all the homogeneous polynomials of degree $L$ contained in the ideal $I$. This yields the following result:
\begin{equation}
 \begin{tabular}{|l|ccccccccc|}
\hline
$L$ & 0 & 1 & 2 & 3 & 4 & 5 & 6 & 7 & 8 \\
\hline
$\dim R(\CPP^5)_L$ & 1 & 6 & 21 & 56 & 126 & 252 & 462 & 792 & 1287 \\
$\dim I_L$ & 0 & 0 & 6 & 28 & 81 & 186 & 371 & 672 & 1134\\
$\dim \tilde{\CR}_L$& 1 & 6 & 15 & 28 & 45 & 66 & 91 & 120 & 153 \\
$\dim \tilde{\CR}_{2L-1}$ for $\CPP^2$ & 1 & 6 & 15 & 28 & 45 & 66 & 91 & 120 & 153 \\
\hline
\end{tabular}
\end{equation}
For comparison, we also listed the dimensions for $\CR_{2L-1}$ in the case of fuzzy $\CPP^2$. Note that the dimensions of the matrix algebras agree; as we will see later, however, the geometry of the spaces which is captured by a Laplace operator acting on these algebras differs. The dimensions of the matrix algebras $\tilde{\CA}_L$ is the evidently the square of $\dim \tilde{\CR}_L$.

\subsection{Embedding of $\CAt_{L-1}$ in $\CAt_{L}$}

As in the case of $\CPP^n$, each algebra at level $L-1$ can be identified with a subalgebra of $\CAt_L$. Since the projectors $\CPh_{L}$ do not commute with the Laplacian $\hat{\Delta}$, this identification is slightly nontrivial. First, note that we have the following natural embeddings:
\begin{equation}
\CAt_{L-1} \ =\ \CA_{L-1}(\VV_{2,2}) \ \embd\ \CA_{L-1}(\CPP^5)\ \embd\ \CA_L(\CPP^5)  \hookleftarrow \CA_L(\VV_{2,2}) \ =\ \CAt_{L}~.
\end{equation}
Furthermore, we have the projectors $\CPh_{L}$ from above, giving a map  $\CA_L(\CPP^5)\rightarrow \CA_L(\VV_{2,2})$.

It is important that the composite map $\CA_{L-1}(\VV_{2,2})\rightarrow \CA_{L}(\VV_{2,2})$ is injective, which is most easily seen using the equivalent description in terms of polynomials. Given two operators $\fh_1,\fh_2$ in $\CAt_{L-1}$, they are mapped to $f_1(w,\bw)$ and $f_2(w,\bw)$ by $\tilde{\CF}_L$. Assume that the operators are not equivalent, $\fh_1\nsim\fh_2$, i.e., that $f_1(w,\bw)-f_2(w,\bw)\notin I$ or, equivalently, $f_1(w(z),\bw(\bz))\neq f_2(w(z),\bw(\bz))$. The operators $\ah_i^\dagger\fh_1\ah_i$ and $\ah_i^\dagger\fh_2\ah_i$ are now mapped to $w_i(z)\bw_i(\bz)f_1(w(z),\bw(\bz))$ and $w_i(z)\bw_i(\bz)f_2(w(z),\bw(\bz))$, which cannot be equal and therefore two in-equivalent operators in $\CAt_{L-1}$ are mapped to two in-equivalent operators in $\CAt_{L}$. In other words, we make an error by projecting after multiplying the monomials of order $L-1$ by $w_i\bw_i$, but this error is proportional to terms in $I_L\otimes R_L^*\cup R_L\otimes I_L^*$ and thus vanishes on the Veronese surface.

Note that the matrix algebra $\CAt_L$ is isomorphic to the matrix algebra of $\CPP^2_F$ at level $2L$. This follows immediately by recalling that we consider only operators which map to in-equivalent functions under $\tilde{\CF}_L:\CA_L(\VV_{2,2})\rightarrow \CM_{2L}(\CPP^2)$. At level 2, e.g., there are $21$ monomials spanning $R(\CPP^5)_2$ and $15$ monomials spanning $R(\CPP^2)_4$; the number of monomials spanning $(R(\CPP^5)/I)_2$ is also $21-6=15$. For embedding $\CA_{L-1}(\VV_{2,2})$ into $\CA_L(\VV_{2,2})$ we could therefore have also used the embedding
\begin{equation}
\CA_{L-1}(\VV_{2,2})\ \cong\ \CA_{2L-2}(\CPP^2)\ \embd\ \CA_{2L}(\CPP^2)\ \cong\ \CA_L(\VV_{2,2})~.
\end{equation}
In our above considerations, this would amount to an embedding by multiplying by $w_0\bw_0+w_1\bw_1+w_2\bw_2+2w_3\bw_3+2w_4\bw_4+2w_5\bw_5$ instead of $w_i\bw_i$. The first convention, however, has some advantages as it is compatible with the geometry of $\CPP^5$, which in turn is responsible for classifying the algebras $\CA_L$ and $\CAt_L$. Furthermore, the second convention does not generalize to arbitrary subvarieties of $\CPP^n$.

\subsection{Riemannian geometry}

It remains to provide the geometry of the Veronese surface as additional information to the given matrix algebra. That is, we have to define a Laplace operator, and the way we represented the matrix algebra suggests to take the one obtained by restricting the Laplace operator on $\CPP^5$ to $\VV_{2,2}$. At the same time, it is clear that the projection of the operator algebra from $\CPP^5$ to $\VV_{2,2}$ will not be compatible (i.e.\ it will not commute) with the decomposition of the operator algebra into eigensubspaces of the $\CPP^5$-Laplace operator $\hat{\Delta}$. This is reflected in the additional projection we used in the embedding of $\CA_{L-1}(\VV_{2,2})$ into $\CA_L(\VV_{2,2})$. We are thus led to introduce the representation of the restriction of a derivative $\hat{L}^a$ on $\CPP^5_L$ as
\begin{equation}
\rho(\tilde{\hat{L}}^a)\CPh_L\hat{f}\CPh_L\ :=\ \rho\left(\left.\hat{L}^a\right|_{\VV_{2,2}}\right) \CPh_L \hat{f} \CPh_L\ =\ \CPh_L([\hat{L}^a,\CPh_L \hat{f}\CPh_L])\CPh_L~,
\end{equation}
with $\rho(\cdot)$ being the adjoint action. Therefore, it follows that $\tilde{\hat{L}}^a=\CPh_L\hat{L}^a\CPh_L$. As one easily checks, this derivative satisfies the Leibniz rule. Furthermore, the integral over $\VV_{2,2}$ is defined as $\tr(\CPh_L\,\cdot\,)$ and we have $\tr(\CPh_L [\tilde{\hat{L}}^a\,\cdot\,])=0$, the prerequisite for partial integration.

The definition of the Laplacian is now evident as well:
\begin{equation}
\hat{\Delta}\ :=\ \delta_{ab}\rho(\tilde{\hat{L}}^a)\rho(\tilde{\hat{L}}^b)~.
\end{equation}
This Laplace operator is the one which becomes the ordinary Laplace operator on $\CPP^5$ (and $\VV_{2,2}$) when taken out of the coherent state map:
\begin{equation}\label{TakingOutDelta}
\tilde{\CF}_L(\hat{\Delta}\,\cdot\,)\ =\ \Delta\tilde{\CF}_L(\,\cdot\,)~.
\end{equation}
Just by comparing the spectra of the Laplace operator on $\CPP^2$, which is $k^2+2k$, $k=0,\ldots ,2L$ with the one on $\CPP^5$, which reads as $k^2+5k$, $k=0,\ldots ,L$, we conclude that the metric on both spaces is indeed different. Condition \eqref{TakingOutDelta} actually provides us with a unique Laplace operator on the Veronese surface, even in cases in which the metric on $\VV_{2,2}$ does not descend from the Fubini-Study metric on $\CPP^5$.

Note that at level $1$, the spaces $\VV^F_{2,2}$ and $\CPP^5_F$ are completely indistinguishable since both the matrix algebra and the Laplacian agree. Instead of being disturbing, this feature might give an idea of what interesting properties are to be expected once spacetimes in physics are replaced by more fundamental objects, as e.g.\ fuzzy matrix algebras.

It should be stressed, however, that this definition of a Laplace operator is rather preliminary and only demonstrates the existence of such an object. We will postpone a more detailed analysis of the construction of both Dirac and Laplace operators in the fuzzy case to future work, and focus in the following on the detailed construction of the various matrix algebras corresponding to projective varieties.

\subsection{Alternative description using composite oscillators}

The Veronese varieties -- as well as the Segre varieties and the Gra{\ss}mannians obtained from the Pl{\"u}cker embedding -- allow for an additional description using composite oscillators. That is in the case of $\VV_{2,2}$, we replace
\begin{equation}
\begin{aligned}
(z_0,z_1,z_2) & \ \rightarrow\ (\ah_0,\ah_1,\ah_2)~,\\
(w_0,...w_3) &\ \rightarrow\ (\Ah_0,...,\Ah_5)\ :=\ (\ah_0\ah_0,\ah_1\ah_1,\ah_2\ah_2,\ah_0\ah_1,\ah_0\ah_2,\ah_1\ah_2)
\end{aligned}
\end{equation}
when quantizing the space. The $L$-particle Hilbert space $\tilde{\CR}_L$ obtained from acting with $L$ composite operators $\Ah_i$ on the vacuum is identical to the one obtained from our above construction, and the Laplace operator derived from
\begin{equation}
\tilde{\hat{L}}^a\ :=\ \Ah_i^\dagger \lambda^a_{ij} \Ah_j~,
\end{equation}
where $\lambda^a_{ij}$ are the Gell-Mann matrices of $\asu(6)$, is also equivalent to the one introduced above. More details on using composite oscillators in the construction of fuzzy matrix algebras are found in \cite{Murray:2006pi,Dolan:2006tx}.

\subsection{The most trivial projective subvariety}

The most trivial example of a projective subvariety is in fact the embedding $\CPP^1\embd \CPP^2$ given by 
\begin{equation}
 \CPP^1\ni (z_0,z_1)\mapsto (z_0,z_1,0)=(w_0,w_1,w_2)\in \CPP^2~.
\end{equation}
The ideal to be factored out here is generated by $I_1=w_2$, and accordingly we have the operator $\hat{I}_1=\hat{a}_2$ together with the projector
\begin{equation}
\CPh_{L}:=\unit_L-\frac{1}{\CN_{i_1\ldots i_{L-1}}}\ah_2^\dagger\ah^\dagger_{i_1}\ldots \ah^\dagger_{i_{L-1}}\vac\cav \ah_{i_1}\ldots \ah_{i_{L-1}}\ah_2
\end{equation}
and the resulting algebra of functions $\CAt_L=\CPh_L\CA_L\CPh_L$. The quantization of this subvariety clearly yields the usual fuzzy algebra of functions on $\CPP^1_F$ and it is hence equivalent to the ordinary quantization procedure. Even the Laplace operator obtained on the fuzzy subvariety agrees with the one on $\CPP^1_F$. However, none of the subtle issues in quantizing a projective subvariety appear in this case and we therefore considered the Veronese surface as a first example instead.

\section{Fuzzy weighted projective spaces}

Weighted projective spaces are a first step towards more general fuzzy toric geometries: While the complex projective space $\CPP^n$ is constructed from $\FC^{n+1}\backslash\{0\}$ by factoring out the homogeneous toric action
\begin{equation}
(w_0,\ldots ,w_{n})\sim (\lambda w_0,\ldots ,\lambda w_{n})~,~~~\lambda \in \FC^*,
\end{equation}
the weighted projective spaces $W\CPP^n(p_0,\ldots ,p_{n})$ are obtained from $\FC^{n+1}\backslash\{0\}$ after factoring out the weighted toric action
\begin{equation}
(z_0,\ldots ,z_{m})\sim (\lambda^{p_0}z_0,\ldots ,\lambda^{p_{m}} z_{m})~,~~~\lambda \in \FC^*~.
\end{equation}

\subsection{Embedding of weighted projective spaces in $\CPP^n$}

Consider the weighted projective space $W\CPP^m(p_0,\ldots ,p_{m})$ with all weights $p_i\geq 1$. There are two isomorphisms between weighted projective spaces, which we can use to simplify the discussion \cite{Iano-Fletcher:1989aa}. First, if $q$ is a positive integer, it is 
\begin{equation}
W\CPP^m(p_0,\ldots ,p_m) \ \cong\  W\CPP^m(qp_0,\ldots ,qp_m)~.
\end{equation}
Second, if $(p_0,\ldots ,p_m)$ have no common factor and\footnote{gcd: greatest common divisor or highest common factor} $q=\mathrm{gcd}(p_1,\ldots ,p_m)$, then
\begin{equation}\label{4.4}
W\CPP^n(p_0,\ldots ,p_m) \ \cong\  W\CPP^n(p_0,p_1/q\ldots ,p_m/q)~.
\end{equation}
Thus by repeated application of these isomorphisms, any weighted projective space is isomorphic to one of the form $W\CPP^m(p_0,\ldots ,p_m)$ with $p_0\geq\ldots \geq p_m$ and $\mathrm{gcd}(p_0,\ldots ,\hat{p}_i,\ldots ,p_m)=1$, where the hat $\hat{\cdot}$ indicates an omission.

An embedding into $\CPP^n$ can always be found from a generalization of the Veronese map: take the least common multiple $k$ of the numbers $p_0,\ldots ,p_{m}$ and construct all possible monomials $w_i$ of the coordinates $z_0,\ldots ,z_m$, which transform as $\lambda^k$ by the toric action. The number of the $w_i$ is then equal to $n+1$ and the embedding is given by mapping $(z_0,\ldots ,z_m)$ to $(w_0,...,w_n)$ in an arbitrary order. In general, these monomials will not be independent, and one will arrive at a number of relations $I_i=0$, where $I_i$ are elements of the homogeneous coordinate ring of $\CPP^n$. To arrive at the coordinate ring on the weighted projective space, one has to factor out the ideal generated by the $I_i$.

As an illustration of the procedure, we consider the two examples $W\CPP^2(1,1,2)$ and $W\CPP^2(1,2,2)$. In both cases, the least common multiple is evidently $2$. For the first space, we define an embedding into $\CPP^3$ by
\begin{equation}
W\CPP^2(1,1,2) \ni (z_0,z_1,z_2)\ \mapsto\ (z_0^2,z_1^2,z_0z_1,z_2)\ =\ (w_0,\ldots ,w_3)\in\CPP^3~.
\end{equation}
From the embedding, we can also read off the defining equation $I_1=w_0w_1-w_2^2=0$ for $W\CPP^2(1,1,2)$ in $\CPP^3$. Analogous considerations to the ones outlined in section 3.1 yield straightforwardly that there are no further nontrivial identities than $I_1$, and the homogeneous coordinate ring is just $R/I$, where $R$ is the homogeneous coordinate ring on $\CPP^3$ and $I$ the ideal generated by $w_0w_1-w_2^2$.

The second space turns out to be less interesting, since we are led to the embedding 
\begin{equation}
W\CPP^2(1,2,2)\ni
(z_0,z_1,z_2)\ \mapsto\  (z_0^2,z_1,z_2)\ =\ (w_0,w_1,w_2)\in\CPP^2
\end{equation}
with no defining relation between the coordinates. The coordinate ring on the weighted projective space $W\CPP^2(1,2,2)$ and the one on $\CPP^2$ are thus identical and we conclude that $\CPP^2\cong W\CPP^2(1,2,2)$ in agreement with \eqref{4.4}.
\remark{See Harris, Algebraic Geometry .. a first course, p.127}

\subsection{Fuzzification of weighted projective spaces}

To fuzzify these spaces, we can proceed as in the case of the Veronese surface. That is, we start from operators corresponding to the equations $I_i(z,\bz)=0$, which cut out the weighted projective space from $\CPP^n$. For our example $W\CPP^2(1,1,2)$, we have only one such equation, and the projector $\CPh_{1;L}$ obtained from $I_1$ is therefore given by
\begin{equation}
\CPh_{1;L}\ :=\ \unit_L-\CQh_{1;L}
\end{equation}
at arbitrary level $L$, where
\begin{equation}
\begin{aligned}
\unit_L&\ :=\ \frac{1}{\CN_{i_1\ldots i_L}}\ah^\dagger_{i_1}\ldots \ah^\dagger_{i_{L}}\vac\cav \ah_{i_1}\ldots \ah_{i_{L}}~,\\
\CQh_{1;L}&\ :=\ \frac{1}{\CN_{i_1\ldots i_{L-2}}}\left(\ah_0^\dagger \ah_1^\dagger-\ah_2^\dagger \ah_2^\dagger\right)\ah^\dagger_{i_1}\ldots \ah^\dagger_{i_{L-2}}\vac\cav \ah_{i_1}\ldots \ah_{i_{L-2}}\left(\ah_0 \ah_1-\ah_2 \ah_2\right)~.
\end{aligned}
\end{equation}
More explicitly, we have
\begin{equation}
\begin{aligned}
\hat{\CP}_{1;1}&\ = \ \unit_1\ =\ \hat{a}^\dagger_i\vac\cav\hat{a}_i~,\\
\hat{\CP}_{1;2}&\ = \ \unit_2 - \tfrac{1}{3}\left(\ah_0^\dagger \ah_1^\dagger-\ah_2^\dagger \ah_2^\dagger\right)\vac\cav \left(\ah_0 \ah_1-\ah_2 \ah_2\right)~,\\
\hat{\CP}_{1;3}&\ = \ \unit_3 - \sum_{i=0}^3\frac{1}{\CN_i}\left(\ah_0^\dagger \ah_1^\dagger-\ah_2^\dagger \ah_2^\dagger\right)\hat{a}_i^\dagger\vac\cav\hat{a}_i \left(\ah_0 \ah_1-\ah_2 \ah_2\right)~,
\end{aligned}
\end{equation}
where
\begin{equation}
 \CN_0\ =\ \tfrac{1}{4}~,~~~\CN_1\ =\ \tfrac{1}{4}~,~~~\CN_2\ =\ \tfrac{1}{7}~,~~~\CN_3\ =\ \tfrac{1}{3}~.
\end{equation}

The matrix algebra is obtained as
\begin{equation}
\CAt_L\ :=\ \CPh_{1;L}\CA_L\CPh_{1;L}\ =\ \tilde{\CR}_L\otimes \tilde{\CR}^*_L~,
\end{equation}
and $\tilde{\CR}_L=\CPh_{1;L}\CR_L$ is spanned, e.g.\ at level 2, by
\begin{equation}
\ah_0^\dagger \ah_0^\dagger\vac~,~~~\ah_1^\dagger \ah_1^\dagger\vac~,~~~\left(\ah_0^\dagger \ah_1^\dagger+\tfrac{1}{2}\ah_2^\dagger \ah_2^\dagger\right)\vac~,~~~\ah_0^\dagger \ah_2^\dagger\vac~,~~~\ah_1^\dagger \ah_2^\dagger\vac~,~~~\ah_k^\dagger \ah_3^\dagger\vac~,
\end{equation}
where $k=0,\ldots,3$. These operators are in one-to-one correspondence with the polynomials
\begin{equation}
w_0w_0~,~~~w_1w_1~,~~~w_0w_1+w_2w_2~,~~~w_0w_2~,~~~w_1w_2~,~~~w_k w_3
\end{equation}
spanning $(R/I)_2$.

The dimensions of the matrix algebras $\dim \tilde{\CA}_L=(\dim \tilde{\CR}_L)^2$ are calculated as in the case of the Veronese surface:
\begin{equation}
 \begin{tabular}{|l|ccccccccc|}
\hline
$L$ & 0 & 1 & 2 & 3 & 4 & 5 & 6 & 7 & 8 \\
\hline
$\dim R(\CPP^3)_L$ & 1 & 4 & 10 & 20 & 35 & 56 & 84 & 120 & 165 \\
$\dim I_L$ & 0 & 0 & 1 & 4 & 10 & 20 & 35 & 56 & 84\\
$\dim \tilde{\CR}_L$& 1 & 4 & 9 & 16 & 25 & 36 & 49 & 64 & 81 \\
\hline
\end{tabular}
\end{equation}

Note that the coherent state map $\CF_L:\CA_L\rightarrow \CM_L(\CPP^3)$ gives rise to a bijective map $\tilde{\CF}_L:\CAt_L\rightarrow\tilde{\CM}_L$ by restriction $\tilde{\CF}_L=\CF_L|_{\CAt_L}$. This map defines again a star product. Furthermore, we have an embedding of $\CAt_{L-1}$ in $\CAt_L$ via a similar argument as in the case of the Veronese surface.

\subsection{Singularities in the fuzzy picture}

A new aspect of weighted projective spaces is that -- contrary to the Veronese surfaces -- they are not smooth manifolds in general but contain quotient singularities. 

First, recall that there is a natural notion of a cotangent space on an algebraic variety, which is rather intuitive. The cotangent space is spanned by elements $\dd f$, where $f$ is a linear function and one therefore defines the {\em Zariski cotangent space} of an algebraic variety $X$ at a point $p$ as
\begin{equation}
T^*_p(X)\ :=\  \frm_p/\frm_p^2~,
\end{equation}
where $\frm_p$ is the maximal ideal of functions on $X$ vanishing at $p$. The dimension of $T^*_p(X)$ is in general not constant, and points at which the dimension exceeds the dimension of $X$ are called singular. This directly translates into the following prescription for finding singularities: Given a $d$-dimensional projective algebraic variety $X$ defined in $\CPP^n$ by $k$ algebraic equations $I_i=0$, consider the rank of the matrix
\begin{equation}
J_{ij}\ =\ \left(\derr{I_i}{w_j}\right)_{i=1,\ldots ,k;j=0,\ldots ,n}~,
\end{equation}
where $w_j$ are the homogeneous coordinates on $\CPP^n$, at points $p$ on $X$, i.e.\ $I_i(p)=0$. Wherever the rank of $(J_{ij})$ is smaller than $n-d$, the linearizations of the $I_i$ cut out a tangent space of dimension larger than $d$ and accordingly, the variety has a singular point at $p$.

Let us consider again our example $W\CPP^2(1,1,2)$. The equation embedding this space into $\CPP^3$ is 
$I_1:=w_0w_1-w_2^2=0$. At $w_0=w_1=w_2=0$, the entries of the matrix $J_{1i}$ vanish and thus its rank is $0$. We conclude that $W\CPP^2(1,1,2)$ is singular at the point $(w_0,w_1,w_2,w_3)=(0:0:0:1)$ or $(z_0,z_1,z_2)=(0:0:1)$. 

A natural question to ask at this point is what happens to the singularity under fuzzification. One obviously expects the fuzzy algebra of functions to be insensitive to the singularity. Unfortunately, we do not have a finite matrix algebra on $\CPP^n$ corresponding to holomorphic functions at hand, and thus, we have to switch to the category of real functions. This is easily done by embedding $\CPP^n$ into $\FR^{(n+1)^2}$ using the generators $\lambda^{\ah}_{ij}$ of $\au(n+1)$ as described in section 2.4. For simplicity, let us restrict again to our example $W\CPP^2(1,1,2)$. Switching to real functions, we arrive at a set of equations\footnote{Their explicit form is given in \cite{Balachandran:2001dd}.} defining the embedded $\CPP^3$ as well as additional independent ones corresponding to $w_0w_1-w_2^2=0$ and reducing $\CPP^3$ to $W\CPP^2(1,1,2)$. We can again associate a point $p$ with the ideal of functions vanishing at $p$, and the Zariski cotangent space is defined as above. At the singular point, the dimension of the cotangent space again increases.

In the fuzzy case, we can still find subsets of $\CA_L$ corresponding to operators, which are mapped to sets of (real) functions via $\CF_L$ or $\tilde{\CF}_L$ vanishing at points $p$. However, there is no analogue to the correspondence between points and maximal ideals in the continuum. Since the algebra is noncommutative, we have to distinguish between left- and right-ideals. The only bi-ideals in $\CA_L$ are $0$ and $\CA_L$ itself, as given two elements $\hat{f}, \hat{g}\in \CA_L\backslash\{0\}$, we can always find functions $\hat{h}_1, \hat{h}_2\in \CA_L$ such that $\hat{h}_1\hat{f}\hat{h}_2=\hat{g}$. This statement is obvious from the form of the basis elements \eqref{operatorbasis} of our operator algebra $\CA_L$. Furthermore, in the algebra of real functions, all the operators are hermitian conjugate and therefore a left-ideal is automatically a right-ideal and thus a bi-ideal: Assume $\hat{i}=\hat{i}^\dagger$ generates a left-ideal $I$ and $\hat{f}=\hat{f}^\dagger\in\CA_L$. We then have
\begin{equation}
I\ni\hat{f}\hat{i}\ =\ (\hat{f}\hat{i})^\dagger\ =\ \hat{i}^\dagger\hat{f}^\dagger\ =\ \hat{i}\hat{f}\in I~.
\end{equation}

Altogether, the definition of the Zariski cotangent space breaks down, since we are not able to resolve points in the fuzzy algebra of functions on $\CA_L$ via a correspondence with maximal ideals. The fuzzy picture is therefore necessarily insensitive to singularities.

\section{Toric geometry}

The generalization of our construction to arbitrary complex submanifolds of $\CPP^n$ should by now be obvious. Before we summarize the algorithm for projective toric varieties and present some examples, let us briefly review the construction of these spaces. A more detailed introduction to toric geometry is found e.g.\ in \cite{Fulton:1993aa}, or, more concisely, in \cite{Greene:1996cy} and \cite{Hori:2003ic}, chapter 7.

\subsection{Toric fans}

Toric geometry is essentially the final step of generalizing $\FC^*$-actions on complex manifolds, and a toric space will always be defined as subset $S$ of $\FC^n\backslash\{0\}$ on which the equivalence relation given by the toric action
\begin{equation}
(z_0\ldots ,z_{n})\sim (\lambda_1^{q_{0,1}}\ldots \lambda_j^{q_{0,j}}z_0,\ldots ,\lambda_1^{q_{n,1}}\ldots \lambda_j^{q_{n,j}}z_{n})~,~~~\lambda_1,\ldots ,\lambda_j\in\FC^*~,
\end{equation}
is factored out. Instead of specifying $S$ and the $(q_{i,j})$ explicitly, one is usually given a so-called {\em toric fan}. This is a diagram from which useful information on the corresponding toric variety can be directly read off.

Consider a lattice $\RZ^r$ and its underlying continuum $\FR^r\cong \RZ^r\otimes \FR$. For our purposes, a {\em cone}\footnote{more precisely: a {\em strongly convex rational polyhedral cone}} $\sigma$ is a subset of $\FR^r$, which can be written as a linear combination of elements in $\RZ^r$ with positive coefficients,
\begin{equation}
\sigma\ =\ \{a_i v_i\,|\,a_i\geq 0\mbox{ and } v_i\in \RZ^r\}~,
\end{equation}
together with the condition that $\sigma\cap (-\sigma)=\varnothing$. A collection of cones $\Sigma$ is called a {\em fan} if the intersection of two cones in $\Sigma$ is a face of each of the two cones and each face of a cone is also an element of $\Sigma$. As examples, consider the following two fans:
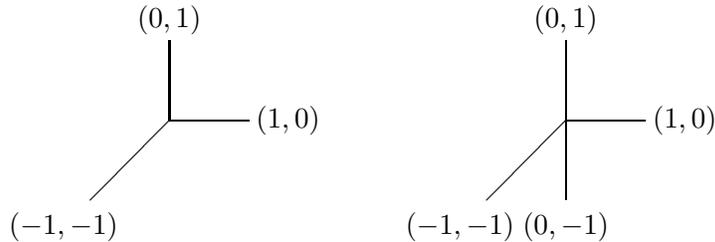
\begin{equation*}
\begin{aligned}
\hspace{5cm}
\begin{picture}(150,80)(0,-35)
\put(0.0,0.0){\line(1,0){30}}
\put(0.0,0.0){\line(0,1){30}}
\put(0.0,0.0){\line(-1,-1){30}}
\put(0.0,38.0){\makebox(0,0)[c]{$(0,1)$}}
\put(45.0,0.0){\makebox(0,0)[c]{$(1,0)$}}
\put(-40.0,-40.0){\makebox(0,0)[c]{$(-1,-1)$}}
\end{picture}&
\begin{picture}(120,80)(0,-35)
\put(0.0,0.0){\line(1,0){30}}
\put(0.0,0.0){\line(0,1){30}}
\put(0.0,0.0){\line(0,-1){30}}
\put(0.0,0.0){\line(-1,-1){30}}
\put(0.0,38.0){\makebox(0,0)[c]{$(0,1)$}}
\put(45.0,0.0){\makebox(0,0)[c]{$(1,0)$}}
\put(-40.0,-40.0){\makebox(0,0)[c]{$(-1,-1)$}}
\put(0.0,-40.0){\makebox(0,0)[c]{$(0,-1)$}}
\end{picture}
\end{aligned}
\end{equation*}
They will turn out to represent the complex projective surface $\CPP^2$ and the Hirzebruch surface $\FF_1$, which is the blow-up of $\CPP^2$ at one point. Identifying the fans with the vectors they are spanned by, we have an (ordered) $n+1$-tuple
\begin{equation}
\Sigma\ =\ (v_0,\ldots ,v_n)~,
\end{equation}
i.e., in the case of our examples,
\begin{equation}
\Sigma_1\ =\ \big((-1,-1),(0,1),(1,0)\big)\eand
\Sigma_2\ =\ \big((-1,-1),(0,1),(1,0),(0,-1)\big)~.
\end{equation}
Define now a map $\phi:\FC^{n+1}\rightarrow \FC^r$ by
\begin{equation}
\phi(t_0,\ldots ,t_n)\ =\ (t_0^{v_0^1}\ldots t_n^{v_n^1},\ldots ,t_0^{v_0^r}\ldots t_n^{v_n^r})~.
\end{equation}
The kernel of $\phi$ consists of the  $(t_0,\ldots ,t_n)$ mapped to $(1,\ldots ,1)$ and defines a toric action on $\FC^{n+1}$ parameterized by elements of $\FC^*$. In the case of our examples, we have
\begin{equation}
\phi_1(t_0,t_1,t_2)\ =\ (t_0^{-1}t_2,t_0^{-1}t_1)\eand
\phi_2(t_0,t_1,t_2,t_3)\ =\ (t_0^{-1}t_2,t_0^{-1}t_1t_3^{-1})~,
\end{equation}
which have nontrivial kernels $(\lambda,\lambda,\lambda)$ and $(\lambda,\lambda\mu,\lambda,\mu)$ and accordingly yield the toric actions
\begin{equation}
(z_0,z_1,z_2)\ \mapsto\  (\lambda z_0,\lambda z_1,\lambda z_2)\eand
(z_0,z_1,z_2,z_3)\ \mapsto\  (\lambda z_0,\lambda\mu z_1,\lambda z_2,\mu z_3)~.
\end{equation}
Evidently, one has to exclude the trivial fixed points of such toric actions from $\FC^{n+1}$, and the prescription for this is as follows: Let $s$ denote a subset of (the vectors spanning) $\Sigma$ which do not form a cone by themselves and construct the linear subspace $V(s)$ by putting the coordinates $z_\alpha$ corresponding to the vectors $v_\alpha\in s$ to zero. The union of all the $V(s)$ forms the set we want to subtract from $\FC^{n+1}$.

Let us now turn to our two examples. In the first case, the only subset $s$ is given by $s=\{v_0,v_1,v_2\}$, and we conclude $V(s)=\{0,0,0\}$. In the second case, we have two such subsets: $\{v_0,v_2\}$ and $\{v_1,v_3\}$, and thus $V(s)=\{0,z_1,0,z_3\}\cup \{z_0,0,z_2,0\}$. Altogether, the first example is indeed the complex projective space $\CPP^2$ and the interpretation of the second space as a blow-up of $\CPP^2$ at one point will become clear in section 7.2.

\subsection{Toric bases}

An alternative description of a projective toric variety is given by so-called {\em integral convex polytopes}, which are convex hulls of lattice points in $\RZ^r$. The idea behind this representation is to factor out any toric action allowed on the variety and arrive at a polytope which forms a ``skeleton'' of the space. Each point on the interior of an $n$-dimensional component of the polytope corresponds to an $n$-dimensional torus; endpoints of edges of the polytope correspond to points on the toric variety, at which the toric action becomes singular. Simple examples of such polytopes are the line segment $[0,1]$ corresponding to the sphere $S^2\cong \CPP^1$ and the triangle with corners $(0,0),(0,1),(1,0)\in\FR^2$, which corresponds to $\CPP^2$. 

Given an integral polytope $\Delta$, one can easily construct the corresponding fan $\Sigma_\Delta$. Consider the inward normals $\vec{n}_F$ on every facet $F$, where a facet is a subset of the polytope with codimension one. Then there are integers $a_F$ such that the polytope is given by 
\begin{equation}
\Delta\ =\ \bigcap_F~\{\vec{p}\in\FR^r\,|\,\langle \vec{p},\vec{n}_F\rangle\geq -a_F\}~.
\end{equation}
The toric fan $\Sigma_\Delta$ corresponding to the polytope $\Delta$ is now spanned by the normals $\vec{n}_F$, and it is a well-known result that the toric variety $X_\Sigma$ is projective precisely if its fan $\Sigma$ originates from an integral polytope. Every facet $F$ of the polytope corresponds to a vector spanning the fan and therefore also to a coordinate $z_F$ on $\FC^{n+1}$. Furthermore, we can associate each integral lattice point $\vec{p}$ in $\Delta$ to a monomial $m(\vec{p})$ according to
\begin{equation}\label{monomialtoric}
m(\vec{p})\ :=\ \prod_F z_F^{\langle \vec{p},\vec{n}_F\rangle+a_F}~.
\end{equation}
Note that the power of each $z_F$ in $m(\vec{p})$ corresponds to the lattice distance from $\vec{p}$ to $F$. All the monomials $m(\vec{p})$ scale with a common factor under arbitrary toric actions and thus provide an embedding of $X_\Delta:=X_{\Sigma_\Delta}$ into $\CPP^{q-1}$, where $q$ is the number of lattice points in $m(\vec{p})$. It is the existence of this embedding which provides the key ingredient for constructing fuzzy toric geometries. As an example, consider the following triple of a polytope, the corresponding normal fan and the monomials $m(\vec{p})$:
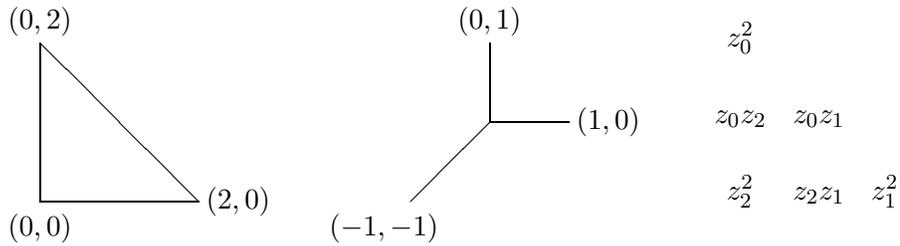
\begin{equation*}
\begin{picture}(170,50)(0,30)
\put(0.0,0.0){\line(2,0){60}}
\put(0.0,0.0){\line(0,2){60}}
\put(0.0,60.0){\line(1,-1){60}}
\put(0.0,68.0){\makebox(0,0)[c]{$(0,2)$}}
\put(75.0,0.0){\makebox(0,0)[c]{$(2,0)$}}
\put(0.0,-10.0){\makebox(0,0)[c]{$(0,0)$}}
\end{picture}
\begin{picture}(80,50)(0,0)
\put(0.0,0.0){\line(1,0){30}}
\put(0.0,0.0){\line(0,1){30}}
\put(0.0,0.0){\line(-1,-1){30}}
\put(0.0,38.0){\makebox(0,0)[c]{$(0,1)$}}
\put(45.0,0.0){\makebox(0,0)[c]{$(1,0)$}}
\put(-40.0,-40.0){\makebox(0,0)[c]{$(-1,-1)$}}
\end{picture}
\begin{array}{ccc} z_0^2 & & \\[0.5cm] z_0z_2 & z_0z_1 & \\[0.5cm] z_2^2 & z_2 z_1 & z_1^2
\end{array}
\vspace{1.0cm}
\end{equation*}
with the vectors $n_0=(0,1)$, $n_1=(1,0)$, $n_2=(-1,-1)$ and the offsets $a_0=a_1=0$ and $a_2=2$. This triple of equivalent data clearly corresponds to the Veronese surface $\CPP^2\embd\CPP^5$. The polytope for $\FF_1$ will be given in section 7.1.

\subsection{Blow-ups}

As in the case of weighted projective spaces, most of the toric varieties will not be smooth but contain singularities. Although one can detect them in the way we analyzed $W\CPP^2(1,1,2)$, there is a more convenient method: Given a toric fan, consider one of its cones. On each ray belonging to this cone, choose the smallest integer lattice point away from the origin. If the simplex obtained from these lattice points has the same volume as the unit simplex in $\FR^n$, then there are no singularities in the corresponding patch. As an example, consider again $W\CPP^2(1,1,2)$ with its toric fan and the derived simplices
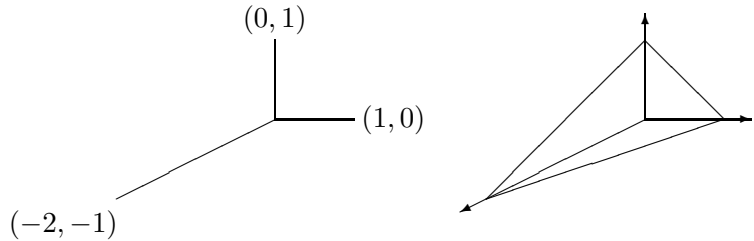
\begin{equation*}
\begin{picture}(80,90)(-40,-40)
\put(0.0,0.0){\line(1,0){30}}
\put(0.0,0.0){\line(0,1){30}}
\put(0.0,0.0){\line(-2,-1){60}}
\put(0.0,38.0){\makebox(0,0)[c]{$(0,1)$}}
\put(45.0,0.0){\makebox(0,0)[c]{$(1,0)$}}
\put(-80.0,-40.0){\makebox(0,0)[c]{$(-2,-1)$}}
\end{picture}
\begin{picture}(100,90)(-100,-40)
\put(0.0,0.0){\vector(1,0){40}}
\put(0.0,0.0){\vector(0,1){40}}
\put(0.0,0.0){\vector(-2,-1){70}}
\put(30.0,0.0){\line(-1,1){30}}
\put(30.0,0.0){\line(-3,-1){90}}
\put(0.0,30.0){\line(-1,-1){60}}
\end{picture}
\end{equation*}
The simplex to the upper left of the origin has twice the volume of the other two, whose volume is that of the unit simplex. The patch at which the variable $z_2$ corresponding to the edge $(1,0)$ is not zero thus contains a singularity, while the other two are non-singular. The only possible singular point is therefore $(z_0,z_1,z_2)=(0:0:1)$.

From this rough analysis of singularities in projective toric varieties, it is clear how to obtain a smooth variety: one needs to subdivide those cones which correspond to singularities until all the simplicial volumes are the ones of the unit simplex. Given an arbitrary algebraic variety containing singularities, one can in fact always perform a finite number of these geometric operations called {\em blow-ups}, which render the variety smooth. Note that a subdivision of a cone in a toric fan corresponds to ``chopping off'' corners in the equivalent toric polytope. In the case of $W\CPP^2(1,1,2)$, e.g., one simply adds the edge $(-1,0)$ and the resulting toric variety is smooth. This geometry is the {\em Hirzebruch surface} $\FF_2$ discussed in section 7.1, and the blow-up (or $\sigma$-process) amounts to replacing the singular point by a $\CPP^1$. 

\section{Fuzzy toric geometries}

\subsection{Fuzzification of a toric variety}

Let us assume that we consider a toric variety $T$ defined by a toric polytope $\Delta$. From this, derive the corresponding toric fan $\Sigma_\Delta$, which leads to a toric action
\begin{equation}
(z_0,\ldots ,z_{n})\ \sim\  (\lambda_1^{q_{0,1}}\ldots \lambda_j^{q_{0,j}}z_0,\ldots ,\lambda_1^{q_{n,1}}\ldots \lambda_j^{q_{n,j}}z_{n})~,~~~\lambda_1,\ldots ,\lambda_j\in\FC^*
\end{equation}
with all the $q_{i,j}$ positive. Each edge in the toric polytope $\Delta$ is associated with a coordinate, and we assign a monomial in the coordinates on $T$ to each integral lattice point within the polytope as discussed in secion 5.2. The number $n+1$ of these monomials tells us, in which $\CPP^n$ the toric variety is embedded and how to build the coordinates $w_i$ on $\CPP^n$ from monomials $w_i=w_i(z_\alpha)$ in the coordinates $z_\alpha$ on $T$. This identification gives rise to a set of identities $I_i=0$, which in turn generate the ideal of (holomorphic) functions, which we wish to factor out. As in sections 3 and 4, we have to make sure that we read off all the $I_i$ using the procedure presented in section 3.1. (Recall furthermore that in the continuum, a homogeneous ideal generated by $I_1$ in $R(\CPP^n)$ can be replace by another homogeneous ideal generated by $\{w_0^kI_1,...,w_n^k I_1\}$, $k\in\NN$, in the definition of a projective subvariety. This will clearly give rise to ambiguities in the quantization process, which are avoided by considering the {\em saturation} of the relevant ideal. Following our recipe, this problem will however not appear.)

To obtain the identities $I_i$, one can use the following simple algorithm. Consider a polytope $\Delta$ with monomials at each integral lattice points and coordinates $z_0,\ldots ,z_r$ associated to the edges of the polytope. A step from one integral lattice point to a neighboring one in a certain direction always changes the powers of the coordinates $z_\alpha$ appearing in the monomials in the same way.
If we identify every monomial at an integral lattice point $\vec{p}$ with an $(r+1)$-dimensional vector $\vec{m}(\vec{p})$ indicating the powers in the coordinates $z_\alpha$ and the change of the powers in the direction $\vec{a}$ with a similar such vector $\vec{\delta}(\vec{a})$, we can write:
\begin{equation}
\vec{m}(\vec{p})+\vec{\delta}(\vec{a})\ =\ \vec{m}(\vec{p}+\vec{a})~.
\end{equation}
Note that multiplying two monomials $m_3=m_1 m_2$ amounts to adding the corresponding vectors $\vec{m}_3=\vec{m}_1+\vec{m_2}$. Consider two partitions $\vec{a}_1+\ldots +\vec{a}_j$ and $\vec{a}'_1+\ldots +\vec{a}'_j$ of a path $\vec{a}$ in the integral lattice and a lattice point $\vec{p}$. Then we have
\begin{equation}
\vec{m}(\vec{p}+\vec{a}_1)+\ldots +\vec{m}(\vec{p}+\vec{a}_j)\ =\ \vec{m}(\vec{p}+\vec{a}'_1)+\ldots +\vec{m}(\vec{p}+\vec{a}'_j)
\end{equation}
and therefore
\begin{equation}
m(\vec{p}+\vec{a}_1)\cdot\ldots\cdot m(\vec{p}+\vec{a}_j)\ =\ m(\vec{p}+\vec{a}'_1)\cdot\ldots\cdot m(\vec{p}+\vec{a}'_j)~.
\end{equation}
By considering all pairs of partitions involving only points inside the polytope, we get all the necessary identities $I_i$. 

As an example, consider the polytope for the Veronese surface $\VV_{2,2}$ with the associated monomials,
\begin{equation}
\begin{array}{ccc}
\vec{p}_0 & & \\
\vec{p}_1 & \vec{p}_2 & \\
\vec{p}_3 & \vec{p}_4 &\vec{p}_5
\end{array}
~~~
\begin{array}{ccc}
z_1^2 & & \\
z_1z_2 & z_0 z_1 & \\
z_2^2 & z_0z_2 & z_0^2
\end{array}
~~~
\begin{array}{ccc}
w_1 & & \\
w_5 & w_3 & \\
w_2 & w_4 & w_0
\end{array}
\end{equation}
From the partitions $\binomr{2}{-2}+\binomr{0}{0}=\binomr{1}{-1}+\binomr{1}{-1}$, $\binomr{2}{0}+\binomr{0}{0}=\binomr{1}{0}+\binomr{1}{0}$ and $\binomr{0}{-2}+\binomr{0}{0}=\binomr{0}{-1}+\binomr{0}{-1}$, we obtain the first three identities 
\begin{equation}
w_0w_1-w_3^2\ =\ 0~,~~~w_0w_2-w_4^2\ =\ 0\eand w_1w_2-w_5^2\ =\ 0~;
\end{equation}
the partitions $\binomr{2}{1}+\binomr{0}{0}=\binomr{1}{0}+\binomr{1}{1}$, $\binomr{1}{-2}+\binomr{0}{0}=\binomr{0}{-1}+\binomr{1}{-1}$ and $\binomr{1}{-1}+\binomr{0}{0}=\binomr{1}{0}+\binomr{0}{-1}$ yield
\begin{equation}
w_3w_4-w_0w_5\ =\ 0~,~~~w_3w_5-w_1w_4\ =\ 0\eand w_4w_5-w_2w_3\ =\ 0~.
\end{equation}

From all of the $I_i$, $i=1,\ldots ,k$, where $k$ is at least\footnote{Recall that e.g.\ in the case of the Veronese surface $\VV_{2,2}$, $k$ was larger.} $n-\dim(T)$, we construct operators $\hat{I}_i$ using the map $\CF_L$ on $\CPP^n$. These operators in turn define projectors $\CQh$ and $\CPh_L$ according to \eqref{defP} and \eqref{defQ}.
The fuzzy algebra of functions $\tilde{\CA}_L$ on $T$ is obtained from the algebra of functions $\CA_L$ on $\CPP^n$ via
\begin{equation}
\tilde{A}_L\ :=\ \CPh_L\CA_L\CPh_L~.
\end{equation}

The operator acting on elements of $\tilde{\CA}_L$ and corresponding to the Laplace operator on $T$ is defined as in the case of the fuzzy Veronese surface and reads as
\begin{equation}
 \hat{\Delta}\ :=\ \delta_{ab}\rho\left(\tilde{\hat{L}}^a\right)\rho\left(\tilde{\hat{L}}^b\right)~,
\end{equation}
where $\rho\left(\tilde{\hat{L}}^a\right)\CPh_L\hat{f}\CPh_L=\CPh_L\left[\hat{L}^a,\CPh_L\hat{f}\CPh_L\right]\CPh_L$ and $\hat{L}^a$ is the generator of $\asu(n+1)$ in the representation acting on the matrix algebra on $\CPP^n$ at level $L$.

\subsection{Fuzzy toric geometry as quantization of the toric base}

Recall that the fuzzy algebra of functions $\CA_L$ on $\CPP^n$ is constructed as the algebra of operators acting on an $L$-particle Hilbert space, which serves as a left-module $\CR_L$. The operators can thus be represented as sums of tensor products of elements of the left-module and the corresponding right-module, $\CA_L\cong \CR_L\otimes \CR_L^*$. For $L=1$, the number of elements in $\CR_L$ equals the number of integral lattice points in the polytope. For higher values of $L$, one might have the idea of replacing the initial integral lattice with one with lattice spacing\footnote{The polytopes obtained in this way are also assigned to the holomorphic line bundles $\CO(L)$ on the complex projective space; see e.g.\ \cite{Hori:2003ic} for details. Furthermore, the relation between $\CO(L)$ and $\CR_L$ is clear from \cite{Dolan:2006tx}.} $1/L$. We will now show that this is indeed what our quantization prescription corresponds to.

Consider a toric variety defined by a polytope with the normals $n_F$ and the offsets $a_F$. The integral lattice points $\vec{p}_i$ in the toric base specified by this data are in one-to-one correspondence with the basis elements $w_i(z_F)$ of $R_1$ and thus also the states spanning $\tilde{\CR}_1$. Multiplying the offsets $a_F$ by $L$ now yields a larger polytope with integral lattice points $\vec{q}_i$. Formula \eqref{monomialtoric} allows us to associate to each lattice point $\vec{q}_i$ various monomials $m(\vec{q}_i)$ in the $w_j(z_F)$, by using partitions of the vectors $\vec{q}_i$ in terms of the vectors $\vec{p}_j$:
\begin{equation}
 m(\vec{q}_i)\ =\ \prod_k w_{j_k}(z_F) \ewith \sum_k \vec{p}_{j_k}\ =\ \vec{q}_i~.
\end{equation}
In the case of $\CPP^n$, there will always be a unique such partition for every $\vec{q}_i$. For more general toric varieties, several partitions might exists and a lattice point might be associated with several monomials. Assigning several monomials to the same lattice point (and averaging over them) corresponds to factoring out the ideal which defines the embedding of the toric variety in a complex projective space. Note that all the monomials are pairwise different when written in terms of the coordinates $z_F$. Thus, there is indeed a one-to-one correspondence between elements of $\tilde{\CR}_L$ and integral lattice points in the enlarged toric polytope. The quantized algebra of functions on a toric variety is therefore obtained by quantizing its toric base.

This procedure allows us to construct the underlying matrix algebra without resorting to the rather inconvenient projectors introduced in section 3. Note, however, that our definition of the Laplace operator given in section 3.5 still requires them. We will look at some explicit examples in the following.

\subsection{Singularities and blow-ups}

One might be tempted to assume that the resolution of singularities which is obtained by rendering the algebra of functions on a singular toric variety fuzzy is in some way connected to a blow-up. This is not so, as a blow-up will always change the number of integral vertices in a toric polytope and thus the fuzzification of a singular toric variety happens in a different complex projective space than the fuzzification of its blow-up. However, there are often several ways of rendering a singular projective toric variety smooth and they yield varieties, which are related via so-called {\em flop transitions}. The meaning of such transitions in the fuzzy picture certainly deserves further study.

\section{Examples}

In this section, we briefly present some of the probably most interesting fuzzy toric geometries. That is, we will make explicit the fuzzification of some projective toric varieties up to the point where expressions for the operators $\hat{I}_i$ are found, from which the derivation of corresponding projectors $\CPh_L$ and thus the construction of the fuzzy algebra of functions is straightforward. If ones interest in fuzzy geometry is essentially coming from regularizing four-dimensional quantum field theories, the most interesting toric geometries are certainly complex surfaces of which the most prominent candidates\footnote{A complete classification of compact complex surfaces is given by the Enriques-Kodaira classification.} are probably the Hirzebruch surfaces $\FF_n$ and the del Pezzo surfaces $\DD_d$, see e.g.\ \cite{Friedman:1998aa} for further details. Interestingly, these two species also play an important r{\^o}le in string compactification, which in turn naturally leads to Calabi-Yau manifolds embedded in complex projective spaces.

\subsection{Fuzzy Hirzebruch surfaces}

Hirzebruch surfaces are particularly interesting as together with $\CPP^2$, they provide a skeleton for smooth rational surfaces in the sense that every such surface can be obtained by a sequence of blow-ups of a Hirzebruch surface $\FF_n$, $n=0$ or $n\geq 2$, or $\CPP^2$. They are therefore called {\em minimal surfaces}. The Hirzebruch surface $\FF_n$ is defined by the toric fan
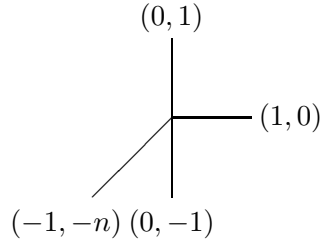
\begin{equation*}
\begin{picture}(10,90)(0,-40)
\put(0.0,0.0){\line(1,0){30}}
\put(0.0,0.0){\line(0,1){30}}
\put(0.0,0.0){\line(0,-1){30}}
\put(0.0,0.0){\line(-1,-1){30}}
\put(0.0,38.0){\makebox(0,0)[c]{$(0,1)$}}
\put(45.0,0.0){\makebox(0,0)[c]{$(1,0)$}}
\put(-40.0,-40.0){\makebox(0,0)[c]{$(-1,-n)$}}
\put(0.0,-40.0){\makebox(0,0)[c]{$(0,-1)$}}
\end{picture}
\end{equation*}
where we order the edges according to $\big((1,0),(-1,-n),(0,-1),(0,1)\big)$.
This fan yields an embedding of $\FF_n$ in $\FC^2\backslash\{0\} \times\FC^2\backslash\{0\}$ with the identification
\begin{equation}
(z_0,z_1,z_2,z_3)\ \sim\ (\lambda z_0,\lambda z_1,\lambda^n \mu z_2,\mu z_3)~,~~~\lambda,\mu \in \FC^*~. 
\end{equation}
Alternatively, one can write $\FF_n=\PP(\CO(n)\oplus \CO(0))$ and accordingly, $\FF_n$ is a $\CPP^1$-bundle over $\CPP^1$. We have $\FF_0=\CPP^1\times \CPP^1$ and $\FF_1=\sigma_p \CPP^2$. Although there is an embedding $\FF_1\embd \CPP^4$ obtained in the usual way from the polytope corresponding to the toric fan, let us give an embedding into $\CPP^2\times \CPP^1$, which shows that $\FF_1$ is indeed a blow-up at one point:
\begin{equation}
\FF_1\ \cong\ \left\{((x_0,x_1,x_2),(y_0,y_1))\in \CPP^2\times \CPP^1 ~|~ x_1y_0=x_0y_1 \right\}~.
\end{equation}
The relation with the homogeneous coordinates $(z_\alpha)$ is given by
\begin{equation}
x_0\ =\ z_0z_3~,~~~x_1\ =\ z_1z_3~,~~~x_2\ =\ z_2~,~~~y_0\ =\ z_0~,~~~y_1\ =\ z_1~,
\end{equation}
which is bijective and guarantees $x_1y_0=x_0y_1$. Note that at every point except for the point $p=(0:0:1)$, $\FF_1$ is identical to $\CPP^2$, as the coordinates on $\CPP^1$ are fixed by the constraint. At $p$, however, the constraint is automatically satisfied, and we gain the freedom to specify a point on a sphere.

In the case of general $\FF_n$, the polytope corresponding to the above fan gives rise to the following monomial structure:
\begin{equation}
\begin{array}{ccccc}
w_0\ =\ z_0^{n+1}z_3 & w_1\ =\ z_0^nz_1z_3 & \ldots  & w_{n}\ =\ z_0z_1^{n}z_3 & w_{n+1}\ =\ z_1^{n+1}z_3 \\
&&&w_{n+2}\ =\ z_0z_2 & w_{n+3}\ =\ z_1z_2
\end{array}~.
\end{equation}
This implies that the fuzzification of $\FF_n$ happens in $\CPP^{3+n}$. The identities are found from two kinds of paths in the polytope: purely horizontal ones and those containing a step in the negative vertical direction. Again, by the algorithm proposed in section 3.1, one can prove that all the nontrivial identities are the ones at $L=2$, i.e.\ those involving products of two monomials. The first kind of paths gives rise to the operators
\begin{equation}
\hat{I}_{ijk}\ =\ \ah_i\ah_j-\ah_{i+k}\ah_{j-k}\ewith i\ <\ j-1,~k\ <\ j-i~,
\end{equation}
while the second kind of paths yields
\begin{equation}
\hat{I}_i\ =\ \ah_{n+2}\ah_{i+1}-\ah_{n+3}\ah_i~,~~~i\ =\ 0,\ldots,n~.
\end{equation}
From these operators, the projectors $\CPh_L$ are easily constructed, and the fuzzy algebra of functions reads as $\CAt_L=\CPh_L\CA_L\CPh_L$, where $\CA_L$ is the algebra of fuzzy functions on $\CPP^{3+n}$ at level $L$. 

As an example, let us look in more detail at the Hirzebruch surface $\FF_1$. For $L=1$ we obtain the toric polytope from the normals $n_0=(-1,0)$, $n_1=(1,1)$, $n_2=(0,-1)$, $n_3=(0,1)$ together with the offsets $a_0=0$, $a_1=2$, $a_2=0$, $a_3=1$. We then have the following integral lattice points in the toric base of $\FF_1$:
\begin{equation}
\begin{array}{ccc}
 w_0=z_0^2z_3 & w_1=z_0z_1z_3 & w_2=z_1^2z_3\\
 & w_3=z_0z_2 & w_4=z_1 z_2
\end{array} 
\end{equation}
The operators corresponding to the ideal read here as
\begin{equation}
 \hat{I}_{021}\ =\ \ah_0\ah_2-\ah_1^2~,~~~\hat{I}_0\ =\ \ah_3\ah_2-\ah_4\ah_0\eand\hat{I}_1\ =\ \ah_3\ah_2-\ah_4\ah_1~.
\end{equation}
We now enlarge $a_F$ by replacing it with $2a_F$. The resulting polytope contains integral lattice points with associated monomials 
\begin{equation}
 \begin{array}{ccccc}
  z_0^4z_3 & z_0^3z_1z_3^2 & z_0^2z_1^2z_3^2 & z_0z_1^3z_3^2 & z_1^4z_3^2 \\ & z_0^3z_2z_3 & z_0^2z_1z_2z_3 & z_0z_1^2z_2z_3 & z_1^3z_2z_3 \\ & & z_0^2z_2^2 & z_0z_1z_2^2 & z_1^2z_2^2
 \end{array}
\end{equation}
or, in terms of the coordinates $w_i$:
\begin{equation}
 \begin{array}{ccccc}
w_0^2 & w_0w_1 & w_1^2=w_0w_2 & w_1w_2 & w_2^2 \\
& w_0w_3 & w_1w_3=w_0w_4 & w_1w_4=w_3w_2 & w_2w_4 \\
& & w_3^2 & w_3w_4 & w_4^2
 \end{array}
\end{equation}
from which we read off the following basis of $\tilde{\CR}_2$:
\begin{equation}
 \begin{array}{ccccc}
(\ah_0^\dagger)^2\vac & \ah^\dagger_0a^\dagger_1\vac & \left(\tfrac{1}{2}(\ah^\dagger_1)^2+\ah^\dagger_0\ah^\dagger_2\right)\vac & \ah^\dagger_1\ah^\dagger_2\vac & (\ah^\dagger_2)^2\vac \\
& \ah^\dagger_0\ah^\dagger_3\vac & \left(\ah^\dagger_1\ah^\dagger_3+\ah^\dagger_0\ah^\dagger_4\right)\vac & \left(\ah^\dagger_1\ah^\dagger_4+\ah^\dagger_3\ah^\dagger_2\right)\vac & \ah^\dagger_2\ah^\dagger_4\vac \\
& & (\ah^\dagger_3)^2\vac & \ah^\dagger_3\ah^\dagger_4\vac & (\ah^\dagger_4)^2\vac
 \end{array}
\end{equation}
For more general $L$, we can easily establish the formula 
\begin{equation}
 \dim\tilde{\CR}_L\ =\ 1+L+\tfrac{3}{2}L(1+L)~,
\end{equation}
from which the dimension of the matrix algebra $\tilde{\CA}_L$ is found via $\dim \tilde{\CA}_L=(\dim \tilde{\CR}_L)^2$.

\subsection{Fuzzy del Pezzo surfaces}

Another class of surfaces are the del Pezzo surfaces $\DD_d$, which are complex two-dimensional Fano varieties. In general, they are the blow-up of $9-d$ generic points on $\CPP^2$. There is a subset of toric del Pezzo surface, $\DD_9\cong\CPP^2$, $\DD_8':=\FF_0$, $\DD_8=\FF_1$, $\DD_7$ and $\DD_6$, whose fans look like
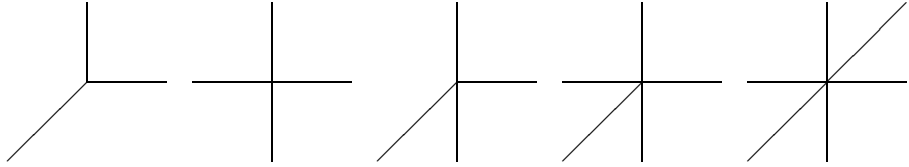
\begin{equation*}\hspace{2cm}
\begin{picture}(70,90)(0,-40)
\put(0.0,0.0){\line(1,0){30}}
\put(0.0,0.0){\line(0,1){30}}
\put(0.0,0.0){\line(-1,-1){30}}
\end{picture}
\begin{picture}(70,90)(0,-40)
\put(0.0,0.0){\line(1,0){30}}
\put(0.0,0.0){\line(0,1){30}}
\put(0.0,0.0){\line(0,-1){30}}
\put(0.0,0.0){\line(-1,0){30}}
\end{picture}
\begin{picture}(70,90)(0,-40)
\put(0.0,0.0){\line(1,0){30}}
\put(0.0,0.0){\line(0,1){30}}
\put(0.0,0.0){\line(0,-1){30}}
\put(0.0,0.0){\line(-1,-1){30}}
\end{picture}
\begin{picture}(70,90)(0,-40)
\put(0.0,0.0){\line(1,0){30}}
\put(0.0,0.0){\line(0,1){30}}
\put(0.0,0.0){\line(0,-1){30}}
\put(0.0,0.0){\line(-1,0){30}}
\put(0.0,0.0){\line(-1,-1){30}}
\end{picture}
\begin{picture}(70,90)(0,-40)
\put(0.0,0.0){\line(1,0){30}}
\put(0.0,0.0){\line(0,1){30}}
\put(0.0,0.0){\line(0,-1){30}}
\put(0.0,0.0){\line(-1,0){30}}
\put(0.0,0.0){\line(-1,-1){30}}
\put(0.0,0.0){\line(1,1){30}}
\end{picture}
\end{equation*}
where all the endpoints are of the form $(a,b)$ with $a,b\in \{-1,0,1\}$.

From the above discussion, it is clear how to render the algebras of functions on\footnote{The del Pezzo surface $\DD_9$ is more precisely the Veronese surface $\VV_{2,3}$, for which the quantization is also clear.} $\DD_9$, $\DD_8'$ and $\DD_8$ fuzzy. For $\DD_7$, assign coordinates $(z_0,\ldots ,z_4)$ to the edges in the toric fan in a counter-clockwise direction, starting at $(1,0)$. Then the equivalence relation reads as
\begin{equation}
(z_0,z_1,z_2,z_3,z_4)\sim(\lambda \mu z_0,\lambda \kappa z_1,\mu z_2,\lambda z_3,\kappa z_4)~,~~~\kappa,\lambda,\mu\in \FC^*~,
\end{equation}
and we arrive at a polytope given by the normals $n_0=(-1,0)$, $n_1=(0,-1)$, $n_2=(1,0)$, $n_3=(1,1)$, $n_4=(0,1)$ and the offsets $a_0=0$, $a_1=0$, $a_2=2$, $a_3=3$, $a_4=2$. We attach the following monomials to the integral lattice points contained in this polytope:
\begin{equation}
\begin{array}{ccc}
w_0\ =\ z_0^2z_3z_4^2 & w_1\ =\ z_0z_2z_3^2z_4^2 & w_2\ =\ z_2^2z_3^3z_4^2\\
w_3\ =\ z_0^2z_1z_4 & w_4\ =\ z_0z_1z_2z_3z_4 & w_5\ =\ z_1z_2^2z_3^2z_4 \\
&w_6\ =\ z_0z_1^2z_2 & w_7\ =\ z_1^2z_2^2z_3
\end{array}
\end{equation}
The quantization of $\DD_7$ thus makes use of the fuzzy algebra corresponding to\footnote{Note that more generally, one should consider the del Pezzo surface of degree $d$ as a subvariety of $\CPP^d$.} $\CPP^7_F$. As far as the relevant identities are concerned, we obtain from the purely horizontal and purely vertical paths $\binomr{2}{0}$ and $\binomr{0}{-2}$ the following operators at level 2:
\begin{equation}
\hat{I}_1\ =\ \ah_0\ah_2-\ah_1^2~,~~~\hat{I}_2\ =\ \ah_3\ah_5-\ah_4^2~,~~~
\hat{I}_3\ =\ \ah_1\ah_6-\ah_4^2~,~~~\hat{I}_4\ =\ \ah_2\ah_7-\ah_5^2~.
\end{equation}
We have furthermore
\begin{equation*}
 \begin{aligned}
  \binomr{1}{-1}\,:&~~~ &&\hat{I}_5\ =\ \ah_0\ah_4-\ah_1\ah_3~,~~~&&\hat{I}_6\ =\ \ah_1\ah_5-\ah_2\ah_4~,~~~&&\hat{I}_7\ =\ \ah_4\ah_7-\ah_5\ah_6~,~~~\\
  \binomr{2}{-1}\,:&~~~ &&\hat{I}_8\ =\ \ah_0\ah_5-\ah_2\ah_3~,~~~&&\hat{I}_9\ =\ \ah_0\ah_5-\ah_1\ah_4~,~~~&&\hat{I}_{10}\ =\ \ah_3\ah_7-\ah_4\ah_6~,~~~\\
  \binomr{1}{-2}\,:&~~~ &&\hat{I}_{11}\ =\ \ah_0\ah_6-\ah_3\ah_4~,~~~&&\hat{I}_{12}\ =\ \ah_1\ah_7-\ah_6\ah_2~,~~~&&\hat{I}_{13}\ =\ \ah_1\ah_7-\ah_4\ah_5~,~~~\\
\binomr{2}{-2}\,:&~~~ &&\hat{I}_{14}\ =\ \ah_0\ah_7-\ah_4\ah_4~,
 \end{aligned}
\end{equation*}
and the prescription of section 3.1.\ shows after some work that there are no further nontrivial identities at higher levels. The enlarged toric polytope with $a_F\rightarrow 2 a_F$ contains integral lattice points with the following monomials in terms of the $w_i$ attached to them:
\begin{equation*}
 \begin{array}{ccccc}
  w_0^2 & w_0w_1 & w_1^2=w_0w_2 & w_1w_2 & w_2^2\\
w_0w_3 & w_0w_4=w_1w_3 & w_0w_5=w_1w_4=w_3w_2 & w_1w_5=w_2w_4 & w_2w_5\\
w_3^2&w_3w_4=w_0w_6 & w_4^2=w_1w_6=w_3w_5=w_0w_7 & w_4w_5=w_1w_7=w_2w_6 & w_5^2=w_2w_7\\
&w_3w_6 & w_4w_6=w_3w_7 & w_4w_7=w_5w_6 & w_5w_7\\
&&w_6^2 & w_6w_7 & w_7^2
 \end{array}
\end{equation*}
This gives rise to the following basis of $\tilde{\CR}_2$:
\begin{equation*}
\begin{aligned} 
 &(\hat{a}^\dagger_0)^2\vac~,~~~\hat{a}^\dagger_0\hat{a}^\dagger_1\vac~,~~~\left(\tfrac{1}{2}(\hat{a}^\dagger_1)^2+\hat{a}^\dagger_0\hat{a}^\dagger_2\right)\vac~,~~~\hat{a}^\dagger_1\hat{a}^\dagger_2\vac~,~~~(\hat{a}^\dagger_2)^2\vac~,~~~
\hat{a}^\dagger_0\hat{a}^\dagger_3\vac~,\\
&\left(\hat{a}^\dagger_0\hat{a}^\dagger_4+\hat{a}^\dagger_1\hat{a}^\dagger_3\right)\vac~,~~~\left(\hat{a}^\dagger_0\hat{a}^\dagger_5+\hat{a}^\dagger_1\hat{a}^\dagger_4+\hat{a}^\dagger_3\hat{a}^\dagger_2\right)\vac~,~~~\left(\hat{a}^\dagger_1\hat{a}^\dagger_5+\hat{a}^\dagger_2\hat{a}^\dagger_4\right)\vac~,~~~\hat{a}^\dagger_2\hat{a}^\dagger_5\vac~,~~~
\\&(\hat{a}^\dagger_3)^2\vac~,~~~\left(\hat{a}^\dagger_3\hat{a}^\dagger_4+\hat{a}^\dagger_0\hat{a}^\dagger_6\right)\vac~,~~~\left(\tfrac{1}{2}(\hat{a}^\dagger_4)^2+\hat{a}^\dagger_1\hat{a}^\dagger_6+\hat{a}^\dagger_3\hat{a}^\dagger_5+\hat{a}^\dagger_0\hat{a}^\dagger_7\right)\vac~,~~~\\&\left(\hat{a}^\dagger_4\hat{a}^\dagger_5+\hat{a}^\dagger_1\hat{a}^\dagger_7+\hat{a}^\dagger_2\hat{a}^\dagger_6\right)\vac~,~~~\left(\tfrac{1}{2}(\hat{a}^\dagger_5)^2+\hat{a}^\dagger_2\hat{a}^\dagger_7\right)\vac~,~~~
~,~~~\hat{a}^\dagger_3\hat{a}^\dagger_6\vac~,\\&\left(\hat{a}^\dagger_4\hat{a}^\dagger_6+\hat{a}^\dagger_3\hat{a}^\dagger_7\right)\vac~,~~~\left(\hat{a}^\dagger_4\hat{a}^\dagger_7+\hat{a}^\dagger_5\hat{a}^\dagger_6\right)\vac~,~~~\hat{a}^\dagger_5\hat{a}^\dagger_7\vac~,~~~
(\hat{a}^\dagger_6)^2\vac~,~~~\hat{a}^\dagger_6\hat{a}^\dagger_7\vac~,~~~(\hat{a}^\dagger_7)^2
\end{aligned}
\end{equation*}
More generally, we have 
\begin{equation}
\dim(\tilde{\CR}_L)\ =\ 1+\tfrac{7}{2}L(1+L)~.
\end{equation}

In the case of $\DD_6$, we assign coordinates to the edges of the toric fan as above and arrive at the equivalence relation
\begin{equation}
(z_0,z_1,z_2,z_3,z_4,z_5)\ \sim\ (\lambda\nu z_0,\mu z_1,\kappa\nu z_2,\lambda z_3,\mu\nu z_4,\kappa z_5)~,~~~\kappa,\lambda,\mu,\nu\in\FC^*~.
\end{equation}
The toric polytope is specified by the normals $n_0=(-1,0)$, $n_1=(-1,-1)$, $n_2=(0,-1)$, $n_3=(1,0)$, $n_4=(1,1)$, $n_5=(0,1)$ together with the offsets $a_0=0$, $a_1=1$, $a_2=0$, $a_3=2$, $a_4=3$, $a_5=2$. This polytope contains integral lattice points with associated monomials
\begin{equation}
\begin{array}{ccc}
w_0\ =\ z_0^2z_1z_4z_5^2 & w_1\ =\ z_0z_3z_4^2z_5^2 &\\
w_2\ =\ z_0^2z_1^2z_2z_5 & w_3\ =\ z_0z_1z_2z_3z_4z_5 & w_4\ =\ z_2z_3^2z_4^2z_5 \\
&w_5\ =\ z_0z_1^2z_2^2z_3 & w_6\ =\ z_1z_2^2z_3^2z_4
\end{array}
\end{equation}
The various paths on the integral lattice lead to the following operators at level 2:
\begin{equation}
 \begin{aligned}
  \binomr{2}{0},~\binomr{0}{-2}\,:&~~~ \hat{I}_1\ =\ \ah_2\ah_4-\ah_3\ah_3~,~~~\hat{I}_2\ =\ \ah_1\ah_5-\ah_3\ah_3~,\\
  \binomr{1}{-1}\,:&~~~ \hat{I}_3\ =\ \ah_0\ah_3-\ah_1\ah_2~,~~~\hat{I}_4\ =\ \ah_3\ah_6-\ah_4\ah_5~,\\
  \binomr{2}{-1}\,:&~~~ \hat{I}_5\ =\ \ah_0\ah_4-\ah_1\ah_3~,~~~\hat{I}_6\ =\ \ah_2\ah_6-\ah_3\ah_5~,\\
\binomr{1}{-2}\,:&~~~ \hat{I}_7\ =\ \ah_0\ah_5-\ah_2\ah_3~,~~~\hat{I}_8\ =\ \ah_1\ah_6-\ah_3\ah_4~,\\
  \binomr{2}{-2}\,:&~~~ \hat{I}_9\ =\ \ah_0\ah_6-\ah_3\ah_3~.
 \end{aligned}
\end{equation}
Enlarging the polytope by replacing $a_F$ by $2a_F$ yields the following monomials:
\begin{equation}
 \begin{array}{ccccc}
  w_0^2 & w_0w_1 & w_1^2 && \\ w_0w_2 & w_1w_2=w_0w_3 & w_1w_3=w_0w_4 & w_1w_4 & \\
  w_2^2 & w_2w_3=w_0w_5 & w_3^2=w_0w_6=w_1w_5=w_2w_4 & w_3w_4=w_1w_6 & w_4^2\\
  & w_2w_5 & w_3w_5=w_2w_6 & w_4w_5=w_3w_6 & w_4w_6\\
&&w_5^2 & w_5w_6 & w_6^2
 \end{array}
\end{equation}
From these, the construction of $\tilde{\CR}_2$ is straightforwardly done as above. The general formula for the dimension of $\tilde{\CR}_L$ reads as
\begin{equation}
 \dim(\tilde{\CR}_L)\ =\ 1+3L(1+L)~.
\end{equation}

\subsection{Fuzzy K3 and fuzzy quintic}

In string theory, most of the interest in toric geometry is not in the toric varieties themselves but in hypersurfaces of these varieties, which contain a large class of Calabi-Yau manifolds \cite{Batyrev-1993}. These manifolds are used in compactifying ten-dimensional superstring theories down to lower dimensions. They admit a Ricci-flat metric in every K{\"a}hler class and have trivial canonical bundle allowing for the volume form to be split into nowhere vanishing holomorphic and antiholomorphic parts. In the case of Calabi-Yau three-folds, one can use the holomorphic part of the volume form to write down an action for a holomorphic Chern-Simons theory \cite{Witten:1992fb}.

A hypersurface in a toric variety $T$ is given as the zero locus of a polynomial $I$ transforming homogeneously under the permitted toric action. In particular, on complex projective spaces $\CPP^n$ with homogeneous coordinates $w_0,\ldots ,w_{n}$, such a polynomial is homogeneous, e.g.
\begin{equation}
 I\ =\ a_0 w_0^d+\ldots +a_{n}w_{n}^d~.
\end{equation}
The degree $d$ of this polynomial is referred to as the degree of the hypersurface.
The conditions for such polynomials to yield hypersurfaces without singularities which are moreover Calabi-Yau are found in \cite{Batyrev-1993}. Two famous examples are the quartic hypersurface in $\CPP^3$ and the quintic hypersurface in $\CPP^4$ giving rise to Calabi-Yau two- and three-folds, the former being called K3 surfaces.

Starting from the algebra of functions at level $L$ on fuzzy $\CPP^3$ and $\CPP^4$, $\CA_{\CPP^3;L}$ and $\CA_{\CPP^4;L}$, the fuzzification of these Calabi-Yau manifolds proceeds precisely as for the embeddings of the projective toric varieties in complex projective space. From the polynomials\footnote{For convenience, we chose the simplest form of the hypersurfaces.}
\begin{equation}
 I_{K3}\ =\ w_0^4+w_1^4+w_2^4+w_3^4\eand
 I_Q\ =\ w_0^5+w_1^5+w_2^5+w_3^5+w_4^5
\end{equation}
one obtains the operators
\begin{equation}
 \hat{I}_{K3}\ =\ \ah_0^4+\ah_1^4+\ah_2^4+\ah_3^4\eand
\hat{I}_Q\ =\ \ah_0^5+\ah_1^5+\ah_2^5+\ah_3^5+\ah_4^5~.
\end{equation}
These in turn give rise to projectors $\CPh_{{K3};L}$ and $\CPh_{Q;L}$ and the algebra of functions on the fuzzy K3 surface and the fuzzy quintic are given by
\begin{equation}
 \CAt_{{K3};L}\ =\ \CPh_{{K3};L}\CA_{\CPP^3;L}\CPh_{{K3};L}\eand\CAt_{Q;L}\ =\ \CPh_{Q;L}\CA_{\CPP^4;L}\CPh_{Q;L}~.
\end{equation}
One should stress that the Laplace operator descending from $\CPP^3$ and $\CPP^4$ to the respective hypersurfaces is not related to a Ricci-flat metric in either cases.

\section{Results and outlook}

In this paper, we showed how to construct fuzzy matrix algebras approximating arbitrary projective toric varieties. We demonstrated that this fuzzification procedure corresponds to a quantization of the toric base. In detail, we discussed the examples of Veronese surfaces, weighted projective spaces, Hirzebruch surfaces, del Pezzo surfaces, K3 surfaces and the quintic in $\CPP^4$. The latter two spaces are Calabi-Yau manifolds, and we thus gain access to the fuzzification of spaces, which not only play an important r{\^o}le in string compactification, but also open perspectives for the definition of holomorphic Chern-Simons theory in the context of fuzzy geometry. Moreover, it seems conceivable that interesting algebraic geometric aspects of theses spaces are reflected in the fuzzy picture, as e.g.\ the relation between different blow-ups of the same singular variety via flop transitions.

It should be stressed that our construction extends to any compact complex manifold, which can be embedded in $\CPP^n$. According to Kodaira's embedding theorem, this holds for any compact complex manifold which admits a positive line bundle or which, equivalently, has a closed positive (1,1)-form $\omega$ whose cohomology class $[\omega]$ is rational.

Besides the  fuzzy flag supermanifolds constructed in \cite{Murray:2006pi}, the description of fuzzy toric geometries creates the possibility of studying mirror symmetry between fuzzy Calabi-Yau manifolds, which in turn might provide further examples on what this symmetry translates to at the level of string geometry, where the fundamental notion of space is no longer that of a manifold.

Beyond these far reaching questions, there are some more immediate open problems arising from our construction. First, one should study in detail the problem of constructing Dirac and Laplace operators for the fuzzy toric geometries. On fuzzy Calabi-Yau manifolds, one might even ask questions about the Laplace operator related to a Ricci-flat metric. Second, it could be interesting to perform numerical studies of scalar models on the various complex surfaces presented in this paper and examine the sensitivity of the model to the different geometries.

One should also stress that the construction of (a majority of) noncommutative vector bundles over the fuzzy toric geometries studied in this paper is rather straightforwardly derived from the one on $\CPP^n_F$, see \cite{Dolan:2006tx}.

Eventually, if physical models on fuzzy superspaces should become important in the study of noncommutative supersymmetric theories or provide a successful regularization for physical theories, one might be interested in extending the constructions obtained here to toric superspaces, as discussed e.g.\ in \cite{Belhaj:2004ts}. This should be rather straightforward, using the construction of fuzzy $\CPP^{m|n}$ as given e.g.\ in \cite{Ivanov:2003qq,Murray:2006pi}.

\newpage
\acknowledgements

I am very grateful to Denjoe O'Connor for suggesting this problem as well as for many fruitful discussions and useful comments on a draft of this paper. Furthermore, I would like to thank Robbert Dijkgraaf, Werner Nahm and Sebastian Uhlmann for helpful discussions. Finally, I gratefully acknowledge financial support from the Dublin Institute for Advanced Studies.

\end{document}